\documentclass[11pt, onecolumn]{IEEEtran}
\usepackage{changepage}
\usepackage{setspace}
\usepackage[english]{babel}
\usepackage[centertags]{amsmath}
\usepackage{amssymb}
\usepackage{amsfonts}
\usepackage{newlfont}
\usepackage{graphicx}
\usepackage{amsmath}
\usepackage{mathtools}
\usepackage{textcomp}
\usepackage[OT1,T1]{fontenc}
\usepackage{authblk}
\usepackage{braket}
\usepackage{booktabs}
\usepackage{geometry}
\usepackage{xcolor}
\usepackage{framed}

\newcommand{\eps}{\varepsilon}

\newcommand{\W}{W_\varepsilon}
\newcommand{\hil}{\mathbb{H}}
\newcommand{\C}{\mathbb{C}}
\newcommand{\m}[1]{\mathcal{#1}}

\newcommand{\RN}[1]{\textup{\uppercase\expandafter{\romannumeral#1}}}

\begin{document}

\title{Randomly Choose an Angle from an Immense Number of Angles
to Rotate Qubits, Compute and Reverse\\[5mm] \Large For QKD Resilient Against Weak Measurements and Securing Entanglement \thanks{We would like to thank the Lynne and William Frankel Center for Computer Science, the Rita Altura Trust Chair in Computer
Science. This research was also partially supported by a grant from the Ministry of Science and Technology, Israel \& the Japan
Science and Technology Agency (JST), and the German Research Funding (DFG, Grant \#8767581199). We also thank Daniel
Berend for discussions, comments and suggestions throughout the research. An extended abstract of this work [BD21] was presented in The Fifth International Symposium on Cyber Security Cryptography
and Machine Learning (CSCML 2021). A preprint version of this paper appears on archive [BD19]. This paper was written when Dor Bitan was a Ph.D. student at the Mathematics department at Ben-Gurion University of the Negev, Beer-Sheva, Israel.}}

\author{\IEEEauthorblockN{Dor Bitan\IEEEauthorrefmark{1} and Shlomi Dolev\IEEEauthorrefmark{2}}\\ \IEEEauthorblockA{\IEEEauthorrefmark{1} Simons Institute for the Theory of Computing, UC Berkeley, Berkeley, USA. \\Email: bitandor@berkeley.edu\\ \IEEEauthorrefmark{2} Dept. of Computer Science, Ben-Gurion University of the Negev, Beer-Sheva, Israel.}} 

\maketitle       
  
\begin{abstract}
\color{black} This paper studies information-theoretically secure quantum homomorphic encryption (QHE) schemes of classical data. Previous works on information-theoretically secure QHE schemes (like Childs'05, Liang'13, and others) are typically based on the Quantum-One-Time-Pad (QOTP) approach of Ambainis et al. [AMTdW'00]. There, the encryption of a bit is a qubit, randomly selected from a set of four possible qubits. This paper takes a different approach and presents the RBE (Random-Basis Encryption) scheme -- a QHE scheme in which the encryption of a bit is a qubit, randomly selected from a set of an immense number of qubits. 

Second, this paper studies weak measurements (WM) and presents a WM-based attack on legacy QOTP-based Quantum Key Distribution (QKD) protocols. Then, we use the RBE scheme to construct a QKD protocol and argue that this protocol is resilient to such WM-based attacks. 

Finally, this paper raises the following question. Entanglement is an essential resource in quantum information and quantum computation research. Hence, once generated, how can its owner secure entangled systems of qubits? We inspect possible QOTP-based solutions, suggest an RBE-based solution, and discuss some of the benefits of the latter.\\

\end{abstract}


\setcounter{page}{1}
\pagenumbering{arabic}
\newpage

\section{Introduction}
\label{sec:4:1}

Delegation of computation, while preserving the confidentiality of the data, is a challenging practical task that has kept researchers busy ever since it was brought up in 1978 by Rivest, Adelman, and Dertouzos \cite{100}. That problem addresses scenarios similar to the following. A user is holding information in the form of a string $x$. The user wishes to use the services of a remote server, which will be referred to as {\it the cloud}, to store $x$ and perform computations over the stored data using computing engines provided by the cloud. In some cases, $x$ is confidential, and hence, the user does not want to share $x$ with the cloud infrastructure enterprises. For example, the user may be a financial company, $x$ is some information regarding the company's financial activity, and the company wishes to use the services of an untrusted cloud to store the data and perform computations over the data.

Existing solutions to the delegation of computation problem are based on either the distributed or the centralized approach. The former typically involves secure multiparty computation protocols and employs several servers (see, e.g., \cite{131} and the references therein),  and the latter often relies on homomorphic encryption (HE) schemes and employs a single server (\cite{132}). When it comes to processing {\it arbitrary} functions over the inputs, both approaches have drawbacks. Indeed, to support the processing of any function over the encrypted data, distributed solutions require ongoing communication between the servers, and centralized solutions can achieve only computational security (and not information-theoretical security). It is noted that computationally secure schemes are based on (a) unproven assumptions regarding the computational hardness of specific mathematical problems, and (b) an assumption that the computing power of the adversary is insufficient for solving instances of these assumed-to-be-hard mathematical problems. On the contrary, IT-secure schemes are free of such assumptions -- their security is derived purely from information theory. This paper seeks IT-secure centralized solutions, and hence is bound to achieve schemes that support only a (proper) subset of all possible functions.

\subsection{Background.}

\noindent {\bf Homomorphic encryption (HE).} Homomorphic encryption (HE) schemes are a core element of centralized solutions to the delegation of computation problem. They may be described by the following collection of four algorithms ($\m{K}, \m{M}$, and $\m{C}$ are the {\it key space}, the {\it message space} and the {\it ciphertext space}, respectively). 

\begin{itemize}
\item \verb|HE.Gen| -- A key generation algorithm which, given a security parameter $n$, outputs a key, $k\in \m{K}$. 
\item \verb|HE.Enc| -- An encryption algorithm which, given a plaintext $x\in\m{M}$ and a key $k\in\m{K}$, outputs a ciphertext $c\in\m{C}$. We will write $c=$\verb|HE.Enc|$_k(x)$ to emphasize that the encryption depends on $k$. 
\item \verb|HE.Eval| -- An evaluation algorithm which, given a function, $f:\m{M} \to \m{M}$ and a ciphertext $c=$\verb|HE.Enc|$_k(x)$, outputs $F(c)$, where $F(c)$ is an encryption of $f(x)$ using the same key. Namely, $F(c)=$\verb|HE.Enc|$_k\bigl(f(x)\bigr)$. 
\item \verb|HE.Dec| -- A decryption algorithm which, given a ciphertext $c=$\verb|HE.Enc|$_k(x)$ and a key, $k$, outputs~$x$.\\
\end{itemize}

\quad The collection satisfies correctness, security, complexity, and other requirements. These requirements come in different flavours, which we now discuss. Conventionally, all algorithms are efficient (i.e., poly-time).
\begin{enumerate}
    \item {\it Security.} Informally, a scheme is secure if the ciphertext leaks a negligible amount of information regarding the plaintext. ``Negligible'' may be interpreted in different ways, and is typically formalized in the IT or computational setting, as discussed above. 
    \item {\it Universality.} The collection of functions that \verb|HE.Eval| supports may be different for different schemes. If \verb|HE.Eval| is defined for all Boolean functions, the scheme is fully homomorphic. As mentioned above, FHE schemes cannot achieve IT-security. IT-secure HE schemes may support different families of functions. 
    \item {\it Compactness.} If \verb|HE.Dec| is efficient (i.e., poly-time), the scheme is {\it compact}. If \verb|HE.Dec| requires $\mathcal{O}(1)$ time and space, the scheme is {\it fully compact}. In some schemes (e.g., most quantum one-time pad based schemes, see below), the evaluation algorithm may output an encryption of the evaluated plaintext that uses a different key. Namely, on input $c=\verb|HE.Enc|_k(x)$, \verb|HE.Eavl| outputs $F(c)=Enc_{k'}(f(x))$, an encryption of $f(x)$ using a different key, $k'\neq k$. Typically, in such schemes, $k'$ depends on $f$, and decryption of the evaluated ciphertext requires the user to modify her keys according to $f$. We stress that such schemes can never achieve full compactness.
    \item {\it Correctness.} Informally, `correctness' means that when decrypting a ciphertext with the right key, the corresponding plaintext is obtained. Namely, a scheme is correct if the two following conditions hold. First, for all $x \in \m{M}$ and $k \in \m{K}$,  $$\verb|HE.Dec|_k\bigl(\verb|HE.Enc|_k(x)\bigr)=x.$$ Second, this is also true if the ciphertext was processed in the cloud. Namely, it holds that for any function $f$ supported by $\verb|HE.Eval|$, for all $x \in \m{M}$ and $k \in \m{K}$, $$\verb|HE.Dec|_k\Bigl(\verb|HE.Eval|_f\bigl(\verb|HE.Enc|_k(x)\bigr)\Bigr)=f(x).$$ A HE scheme that satisfies these two requirements is {\it perfectly correct}. Some HE schemes take a relaxed approach and allow decryption errors with small probability. Namely, they replace the above conditions with the following (relaxed) conditions. First, for all $x \in \m{M}$ and $k \in \m{K}$, $$1 - Pr[\verb|HE.Dec|_k(x)=x]\leq err(n),$$ where $err(\cdot)$ is a function that returns small values for large enough values of $n$. Second, for any function $f$ supported by $\verb|HE.Eval|$, for all $x \in \m{M}$ and $k \in \m{K}$, $$1 - Pr[\verb|HE.Dec|_k\Bigl(\verb|HE.Eval|_f\bigl(\verb|HE.Enc|_k(x)\bigr)\Bigr)=f(x) \leq err(n).$$  
    
    \item {\it Interactivity.} In some HE schemes, the evaluation algorithm $\verb|HE.Eval|$ includes back-and-forth user-server interactions. In other HE schemes, the evaluation algorithm is non-interactive and includes sending but one message from the user to the server with a specification of the function to be evaluated over the ciphertext. \\
\end{enumerate}

\noindent {\bf Quantum computing and quantum homomorphic encryption schemes.} Quantum computers are a threat to computationally secure cryptographic schemes. If built in-scale, quantum computers may be used to efficiently solve problems that are considered intractable for classical computers. Indeed, in 1992 it was shown by Deutsch and Jozsa that quantum computers could solve certain problems exponentially faster than classical computers \cite{102}; Shor suggested in 1994 algorithms that may be invoked by quantum computers to compute discrete logarithms and factor large integers in polynomial time \cite{101}, two problems that are considered computationally hard and stand in the basis of many commonly used computationally secure cryptographic schemes; Grover presented in 1996 a quantum search algorithm that finds a desired record in an $N$ records database in $O(\sqrt{N})$ steps \cite{104}; Bennett and Brassard \cite{129} presented a quantum key distribution (QKD) protocol that enables two distant parties to agree on a random key with IT-security. These are but four celebrated results out of numerous findings established in quantum computation \cite{108}. 

In light of these striking results, a natural question is whether an IT-secure FHE scheme can be built using quantum computers. In 2014, it was shown by \cite{105} that it is impossible to construct an efficient IT-secure quantum FHE (QFHE) scheme. Specifically, the size of the encryption of an IT-secure QFHE scheme grows exponentially with the input size. The non-existence of efficient IT-secure QFHE may also be deduced from different arguments, as in \cite{130}. Either way, efficient IT-secure encryption schemes can be used to homomorphically evaluate only a subset of all possible functions. Such schemes are quantum homomorphic encryption (QHE) schemes, e.g., \cite{128,111,112,113}. Other works use computationally secure FHE schemes to construct computationally secure QFHE schemes. E.g., \cite{121,125,126,127,124}. Quantum schemes with homomorphic properties are often based on the quantum one-time pad (QOTP) encryption scheme, suggested in \cite{122}. There, Pauli gates are randomly applied to the qubits to obtain IT-secure encryption. 

Different schemes are based on different assumptions regarding the capabilities of the parties. QHE schemes typically assume that the server has full quantum capabilities. Assumptions regarding the quantum abilities of the user vary on a broad spectrum between a classical user (with no quantum abilities at all) and a fully quantum user. When the user has (at least some) quantum abilities, the information $x$ held by that user may either be classical or quantum (of course, if the user has no quantum abilities, $x$ can only be classical). In this work, we assume that both the user and the server have full quantum abilities. Namely, they both can: (a) generate qubits in the computational basis; (b) manipulate qubits using quantum logic gates; (c) transmit qubits between each other; (d) measure qubits. We assume that the information held by the user is classical. The function $f$ that is to be homomorphically evaluated over $x$ may either be a classical or quantum algorithm.\\ 

In this work, we look for QHE schemes that enable users to delegate classical data to be stored and processed by an untrusted cloud and have the following properties.
\begin{itemize}
    \item IT-secure.
    \item Efficient. I.e., all algorithms are poly-time.
    \item Fully compact. I.e., the decryption algorithm requires $\mathcal{O}(1)$ time, regardless of $f$. This means that the user is not required to apply any transformations to the encryption keys to decrypt the processed data correctly. 
    \item Perfectly correct. I.e., the ciphertext decrypts to the right plaintext with probability 1 (we ignore errors that may arise due to the nature of noisy physical implementations of quantum schemes). 
    \item Non-interactive. I.e., no client-server interaction is allowed other than the user sending $c=$\verb|HE.Enc|$_k(x)$ to the server, and the server replying with $F(c)=$\verb|HE.Enc|$_k\bigl(f(x)\bigr)$.
\end{itemize} 
 
We ask which operations may be homomorphically applied to encrypted data under these restrictions. Ambianis et al.'s QOTP scheme, suggested in \cite{122}, was used to construct QHE scheme that have some of the properties listed above. Several such schemes are reviewed below. \\

\noindent {\bf Quantum key distribution (QKD) and weak measurements}. In their seminal work from 1984, Bennett and Brassard \cite{129} presented a scheme (hereafter the BB84 protocol) that utilizes a quantum mechanics phenomenon to enable two distant parties, Alice and Bob, to agree on a random key without relying on any computational hardness assumptions. This result indicated that quantum computers could perform tasks that could not be carried using classical computers. 

The BB84 scheme was not only a theoretic breakthrough that paved the way for further theoretic discoveries but was also found to have far-reaching practical applications, as it is feasible to implement it using current-day technology. Private quantum-computing companies (e.g., IDQ and AUREA Technology) offer today quantum-based IT-secure key-exchange services, based on the BB84 protocol, or newer variations of it. Their users include government agencies, financial institutions, companies with distributed offices, and data centers worldwide. 

The security of the BB84 protocol (and the following variations of it) is information-theoretic, i.e., it assumes no limitations on the possible computing power of the adversary. The security of these protocols is based on the laws of quantum mechanics. Mainly, it is based on a postulate of quantum mechanics that states that measurements of a quantum state cause the state to collapse \cite{103}. This phenomenon enables Alice and Bob to reveal eavesdropping attempts. Various attacks on QKD schemes have been suggested over the years. These attacks mainly target weaknesses in the implementation of the scheme and are discussed in, e.g., \cite{142,143,144,145}.

A different approach to attack QKD schemes, which is not based on assumed implementation flaws, and was not previously addressed elsewhere, is based on {\it weak measurements} (hereafter, WM). The model of weak measurements, rooted in the work of Aharonov et al.\ from 1964 \cite{139}, then further developed and studied in, e.g., \cite{140,136,141,135}, raises the possibility of weakly measuring a quantum state. That is, gathering a small amount of information regarding the state while only slightly disturbing it, but not collapsing it. In this work, we investigate ways in which weak measurements incur a threat to the security of QKD schemes. Using weak measurements, an eavesdropper may gather information regarding the key obtained by Alice and Bob, while leaving but slight indications of the eavesdropping that has occurred.

\subsection{Our contribution.} This paper presents four main contributions. We note that, part of the results achieved in this work are briefly described in an extended abstract version of this paper \cite{BD21cscml}, presented in CSCML 2021. These results are brought here fully, along with rigorous proofs, and comparison with previous works. 
\begin{enumerate}
    \item {\it The QHE RBE scheme.} We suggest here a new approach to encrypt and outsource the storage of classical data while enabling limited IT-secure quantum gate computations over the encrypted data. Our method is based on using a specific family of random bases to encrypt classical bits. Our quantum homomorphic encryption (QHE) {\it Random Basis Encryption} (hereafter, RBE) scheme presented here supports fully compact IT-secure homomorphic evaluation of restricted quantum gates over encrypted data. Our RBE scheme is shown to be useful in several applications -- a random basis QKD scheme and a securing entanglement scheme. We note that, while some of these applications may also be constructed using other existing QHE schemes, our scheme has safer security implications in the face of weak measurements. 
    
    Furthermore, in contrast to legacy QHE schemes that require modifications of the keys by the user, our scheme is {\it computation agnostic}. That is, when delegating computations, the user is not required to carry such computations and key-adjustments and can remain utterly oblivious to the implementation method chosen by the cloud.
    
    \item {\it A concrete WM-based attack on two legacy QKD schemes.} We suggest a concrete and novel WM-based attack on two legacy QOTP-based QKD schemes (BB84 and DL04) and formalize and analyze its probability of success. 
    
    \item {\it A novel (RBE-based) QKD scheme -- resilient to our WM-based attacks.} We suggest a novel QKD scheme, and argue that our QKD scheme is resilient to WM-based attacks as those we suggest for the BB84 and DL04 QKD schemes. 
    
    \item {\it Defining the Securing Entanglement problem and suggesting an RBE-based solution.} In this work, we bring a new concept we call {\it securing entanglement}. Entanglement is known to be an essential resource in many quantum settings. The utilization of entanglement in communication, computation, and other scenarios is a very active area of research. In practice, entanglement is usually created by direct interactions between subatomic particles. The creation of entangled systems requires efforts and expenditures. We suggest that, once it was created, this resource should be secured in the sense that only its rightful owners will be able to use it. We demonstrate a process of securing entanglement using our QHE RBE scheme and argue that our method provides safer implications in the face of weak measurements when compared to QOTP based methods. 
\end{enumerate}

Finally, we note that, part of the results presented in this paper, which were recently posted online in a pre-print version of this paper \cite{BD19arxive}, were used as the foundation for the design of quantum multiparty computation schemes (that were also successfully realized on the IBM Quantum Experience platform) \cite{LMHSX20}.

\subsection{Related work.}
\label{sec:4:appBnew}
We now recall recent results in the field and review their attributes with respect to the requirements defined above -- we look for QHE schemes that are IT-secure, efficient, fully-compact, perfectly correct, and non-interactive. Some of these works are also reviewed in a preprint version of this paper \cite{BD19arxive} or in the extended abstract version of this paper \cite{BD21cscml}. These reviews are brought here for completeness.

\noindent (1) {\it Computationally secure QHE schemes}. Broadbent suggested in \cite{120} a client-server scheme based on combining the QOTP encryption scheme with a computationally secure classical FHE scheme. Their scheme enables the delegation of quantum information to a quantum server and homomorphic processing of a universal set of quantum gates over the encrypted data. However, their scheme does not obtain the properties listed above. First, their scheme employs a computationally secure FHE protocol, which makes their scheme only computationally secure (as mentioned, in this work, we are interested in IT-secure schemes). Second, their scheme requires quantum and classical interaction between the user and the server for the processing of non-Clifford gates (while the scope of this work is constructing non-interactive schemes). Third, their scheme is not fully compact, as it requires the user to update the keys used to encrypt the data throughout the computation. Namely, to homomorphically evaluate a quantum circuit over encrypted data, the client must re-adjust her knowledge of the encryption keys on each relevant quantum wire after each gate processing. That re-adjustment requires $\mathcal{O}(s)$ time, where $s$ is the size of the circuit. As mentioned, in this work, we look for fully compact schemes --- schemes in which \verb|HE.Dec| requires $\mathcal{O}(1)$ time. 

An approach similar to \cite{120} was adopted by \cite{121}. There, two schemes were proposed. The first has a decryption procedure whose time-complexity scales with the square of the number of T-gates (and hence does not obtain full compactness). The second scheme uses a quantum evaluation key of length given by a polynomial of degree exponential in the circuit's T-gate depth, yielding a homomorphic scheme only for quantum circuits with constant T-depth. The evaluation key includes auxiliary qubits that encode the required corrections that are to be performed over the processed data. Since a large number of possible corrections must be available, the length of the evaluation key is exponential in the circuit's T-gate depth, yielding a homomorphic scheme that is efficient only for quantum circuits with constant T-depth. Both the schemes of \cite{120} and \cite{121} are only computationally secure (in this work, we are looking for IT-secure schemes).

Dulek et al.\ \cite{125} built on the framework of \cite{121} and used a classical FHE scheme to construct quantum gadgets that allow perfect correction of the errors that occur during the homomorphic evaluation of T-gates on encrypted quantum data. These gadgets give rise to an efficient non-interactive QFHE scheme. Their scheme is compact, but not fully compact since decryption requires the user to apply classical changes to the keys according to $f$. Furthermore, it is only computationally secure.

Mahadev presented in \cite{127} a non-interactive FHE scheme for quantum circuits that is based on QOTP and uses classical keys. The scheme allows a classical user to delegate quantum computations to a quantum server, while the server is unable to learn any information about the computation. Their scheme does not obtain the requirement of perfect correctness as it has positive error probability. Brakerski \cite{124} used the high-level outline of \cite{127} to construct a computationally secure QFHE scheme that enables homomorphic evaluation of classical circuits with bounded depth over classical data and with improved correctness. To support unbounded depth, \cite{127} further rely on a circular security assumption.

The schemes listed above suggest practical solutions to the problem of homomorphic encryption. However, all these schemes have computational security (and not IT-security) and hence does not obtain the properties in which we are interested in this work. The security of their schemes is based on unproven computational hardness assumptions. The schemes listed below rely on no computational hardness assumptions.\\

\noindent (2) {\it Other QHE schemes}. As mentioned above, it was shown in \cite{122} that QOTP is an IT-secure encryption scheme that supports homomorphic evaluation of Pauli gates. Encryption is performed by randomly applying $X$ and $Z$ gates to qubits, conditioned on a two-bit (classical) key, and decryption is performed by applying the same gates in the opposite direction. However, this method alone provides no means for constructing a QHE scheme that withstands our requirements. In particular, homomorphic evaluation of quantum gates over QOTP-encrypted data requires that the user perform computations over the classical keys in compliance with the computations that are performed by the server over the encrypted qubits. This requirement results in decryption complexity linear in the size of the circuit, and hence, the scheme is not fully compact. Set side by side, our scheme is {\it computation agnostic}. That is, the user is not required to carry such key-adjustments, and hence the user may remain agnostic to how the cloud implements the computation. 

Childs \cite{123} discussed ways in which a powerful quantum server may assist a user in performing operations while preserving the confidentiality of the data. In their work, the user is assumed to have capabilities significantly inferior to those of the server. In particular, the user is only allowed to generate qubits in the $\ket{0}$ state, store qubits, perform swap and Pauli gates, and perform no measurements. Under these considerations, they suggest a (QOTP based) way in which the server may perform measurements on encrypted data. They also suggest algorithms that enable the server to help the user perform a universal set of quantum gates over encrypted data. However, these algorithms are neither compact nor non-interactive --- they require the user to perform at least as many operations as the server for each gate, and some of them require rounds of client-server interaction. Hence this scheme is, of course, not computation agnostic.

Rhode et al.\ presented in \cite{128} a protocol that enables a quantum user to manipulate client data in two models of restricted quantum computation --- the boson sampling and quantum walk models. Their protocol is non-interactive, fully compact, and assumes no computational hardness assumptions and no limitations on the computing power of the adversary. However, in their scheme, the same key is used for encoding each of the input qubits, and hence, their scheme withstands no standard cryptographic criterion of security. Tan et al. \cite{112} improved on \cite{128} and presented a protocol that supports a class of quantum computations, including and beyond boson sampling, with improved security (under similar assumptions). However, they achieve no standard criterion of IT-security, as they only bound the amount of information accessible to an adversary.

Ouyang, Tan, and Fitzsimons \cite{113} took a different approach and further improved on the results of \cite{112}. Built on constructions taken from quantum codes, they achieved an encryption scheme that supports the evaluation of circuits with a constant number of non-Clifford gates. Though achieving stronger security guarantees than \cite{128,112}, their scheme withstands no standard cryptographic criterion of security. Furthermore, their scheme is neither perfectly correct nor fully compact. It suggests a tradeoff between the size of the encoding and the success probability, where achieving constant success probability costs in increasing the size of the encoding exponentially with the total number of T gates. 

\cite{111} constructed a QOTP-based quantum encryption scheme which, given the encryption key, permits any unitary transformation to be evaluated on an arbitrary encrypted $n$-qubit state. Their scheme is efficient, compact, and IT-secure against an eavesdropper who may intercept an encrypted message (before or after evaluation). However, their scheme suggests no solution to the main problem discussed in this paper, as their evaluation algorithm is dependent on the key. Under this restriction, the server must hold the key to compute on the encrypted data. Given the key, the server may decrypt and read the message, which by no means provides the user with any level of privacy. They also constructed a scheme in which the evaluation algorithm is independent of the key, but it only supports trivial operations that are independent of the key.\\ 

\noindent (3) {\it Weak measurements and QKD}.
The first QKD scheme was suggested by Bennett and Brassard \cite{129} (hereafter BB84). In BB84, Alice sends Bob a random binary string. Each bit in the string is encoded as a qubit in either the computational basis $\{\ket{0},\ket{1}\}$ or the Hadamard basis $\{\ket{+},\ket{-}\}$. Zero bits ($0$) are encoded by $\ket{0}$ or $\ket{+}$, and one bits ($1$) are encoded by either $\ket{1}$ or $\ket{-}$. In \cite{137}, Deng and Long suggested a two-stage QKD scheme (hereafter DL04). As in \cite{129}, the DL04 protocol uses qubits only in the computational or diagonal basis. This paper argues that using distinct sets of qubits to encode zeros and ones (like in the BB84 and DL04 schemes) is a problematic choice since it may raise security issues when considering Weak Measurement (WM) based attacks. This problematic choice also appears in other QKD schemes. It is noted that in the QKD scheme suggested in this paper, zero bits and one bits can have the same encoding, and hence, the scheme is resilient to weak measurement attacks.

Weak measurements (WM) enable gathering information about the qubit's state while not collapsing it, but only partly biasing it. WM are performed in two steps. First, one weakly interacts the target qubit with an ancilla via a two-qubit gate. Next, the ancilla is (strongly) measured. The (strong) measurement's outcome provides partial information about the target qubit's state. This way, WM allow to outsmart the uncertainty principle.

In \cite{138}, Kak presents a QKD protocol that uses three rounds of communication and a secure quantum channel. There, before executing the protocols, two orthogonal quantum states are chosen as encodings of the bits. Then, a random rotation gate $A$ is applied by Alice to an encoding of her bit $b$, and the bits is transmitted to Bob. Then, a random rotation $B$ is applied by Bob to the qubit and the qubit is transmitted back to Alice, who rotates it in the counter direction by applying $A^\dagger$. Finally, Alice sends the qubit back to Bob, who applies $B^\dagger$ and obtains the encoding of $b$. 

It is noted that Kak's scheme may be resilient to WM-based adversarial attacks. However, the QKD protocol suggested here outperforms Kak's scheme in several aspects. First, the protocol presented here requires only two rounds of interaction, while Kak's scheme requires three -- a 50\% overhead. Second, while in Kak's scheme both Alice and Bob must be able to apply arbitrary quantum gates to qubits, in the QKD protocol presented here only Bob needs to have this ability, and it is enough for Alice to apply only NOT gates. Finally, in Kak's protocol, the parties must choose an encoding of the bits prior to the execution of the protocol. The QKD protocol presented here does not require such a setup stage.

Although previously mentioned in literature, WM were never considered as a plausible technique for attacking QKD protocols. In \cite{147}, an enhanced feedback-control of qubits was demonstrated using weak measurements. In \cite{148}, WM were used in a QKD scheme with an improved key-rate, immunity to detector basis-dependent attacks, and other side-channel attacks. However, WM-based attacks were not considered against the \cite{148} scheme. Their scheme only suggests ways for Alice and Bob to use WM (and not by the adversary). WM were also used in \cite{146} to detect a spin-dependent displacement of photons passing through an air-glass interface. \\

\subsection{Paper organization.} Section \RN{2} provides background on quantum computation. In Section \RN{3}, we present our random basis encryption (RBE) QHE scheme and discuss its homomorphic properties. The concept of securing entanglement is presented in Section \RN{4} and demonstrated through the use of entangled qubits in a pseudo telepathy game. In Section \RN{5}, we describe WM attacks on existing QKD schemes and present our RBE-based WM-resilient QKD scheme. Section \RN{6} concludes the work.

\section{Background on quantum computation}
\label{sec:4:basics}
To address a broad audience, we give a brief overview of the basics of quantum computation. Further details on the topic may be found in \cite{103}, whose notations we follow. Readers who are familiar with the topic can safely skip this section. The basic building block of quantum computation protocols is the {\it qubit}. The qubit is the quantum version of the classical bit used in classical computing. Whereas a classical bit may be described as an element of $\{0,1\}$, a qubit may be described as a unit vector in the Hilbert space $\C^2$. Denote $\hil=\C^2$, and $\ket{0}$ and $\ket{1}$ be the elements $\bigl( \begin{smallmatrix} 1 \\ 0 \end{smallmatrix} \bigr)$ and $\bigl( \begin{smallmatrix} 0 \\ 1 \end{smallmatrix} \bigr)$ of $\hil$, respectively. $\{ \ket{0},\ket{1} \}$ is the {\it computational basis} of $\hil$. We use the Ket notation and denote qubits by $\ket{\psi}$. A system composed of $n$ qubits is described by a unit vector of $\hil^{\otimes n}$, the $n$-fold tensor product of $\hil$ with itself. Such a system of $n$ qubits is the quantum version of an $n$-long string of classical bits. 

An arbitrary qubit $\ket{\psi} \in \hil$ may be described by its coordinates in the computational basis using four real numbers: $\ket{\psi}=\alpha\ket{0}+\beta\ket{1}$, where $\alpha,\beta \in \C$. If $\ket{\psi_1}$ and $\ket{\psi_2}$ are two elements of $\hil$ such that $\ket{\psi_1}= e^{i\gamma}\ket{\psi_2}$ for some $\gamma \in \mathbb{R}$, then $\ket{\psi_1}$ and $\ket{\psi_2}$ are {\it equal up to a global phase factor}. Global phase factors have no influence on quantum computations, and hence may be ignored. Hence, and as $\ket{\psi}$ is a unit vector, one may write $\ket{\psi}$ using only two real numbers: $$\ket{\psi} = \cos(\theta / 2)\ket{0}+e^{i\varphi}\sin(\theta / 2)\ket{1},$$ where $\theta,\varphi\in\mathbb{R}$. This is the {\it Bloch sphere representation} of $\ket{\psi}$. The name ``sphere representation'' comes from the fact that $\theta$ and $\varphi$ may be used to visualize $\ket{\psi}$ as a unit vector in $\mathbb{R}^3$, as described in Figure 1. 

\begin{figure}
\begin{center}
\includegraphics[scale=0.24]{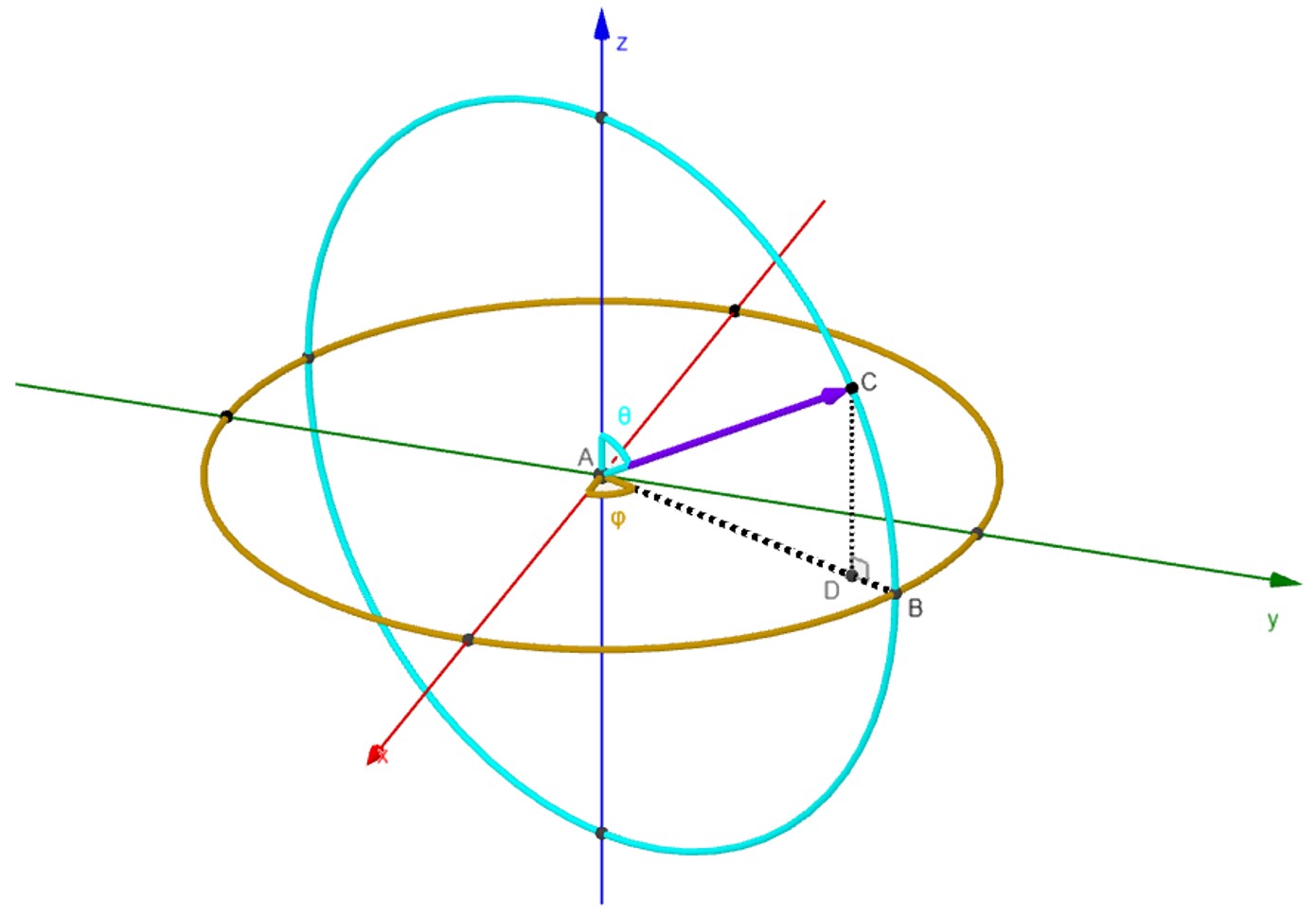}\\
\end{center}

\begin{center}
\begin{minipage}{0.33\linewidth}
\small \textit {Figure 1: Bloch sphere representation.} 
\end{minipage}
\end{center}
\end{figure}

The angle $\varphi$ is taken in the $[xy]$ plane counterclockwise to the positive direction of the $x$ axis and $\theta$ is the angle taken from the positive direction of the $z$ axis in the plain determined by the $z$ axis and the ray determined in the $[xy]$ plain by $\varphi$. $\overrightarrow{AC}$ is the visualization of $\ket{\psi}$ as a unit vector in $\mathbb{R}^3$.\\

In classical computing, strings of classical bits are manipulated using logic gates, information is represented as a string of bits, and the function to be computed over the information is represented as a logic circuit, which is composed of logic gates. In quantum computing, systems of qubits are manipulated using {\it quantum gates}, information is represented as a system of qubits and the function to be computed over the information is represented as a {\it quantum circuit}, which is composed of quantum gates. In order {\it to implement} a classical computation, bits are {\it physically realized} and the physical realizations of the bits are manipulated using physical realizations of logic gates. To implement quantum computations, qubits are physically realized, and these physical realizations of the qubits are manipulated using physical realizations of quantum gates. While classical logic gates are Boolean functions, quantum gates are unitary operators on Hilbert spaces. We use the Kronecker product notation and represent unitary operations as matrices. 

%



Quantum computers may be used to perform computations that have been performed using classical computers, as well as other tasks. For example, any information that may be represented classically as a string of bits may be represented in the quantum model as a tensor product of elements of the computational basis  $\{ \ket{0},\ket{1} \}$ of $\hil$. Then, any classical circuit may be implemented in the quantum model using a quantum circuit composed of {\em Toffoli gates}, which is the quantum version of the classical universal {\it NAND} gate.\\ 


\noindent
{\bf Reading quantum information.} A physical realization of a qubit may come in different forms. However, according to the postulates of quantum mechanics, no matter what form of realization is chosen, given a physical realization of an arbitrary qubit, $\ket{\psi}$, {\it one cannot determine its coordinates.} This phenomenon is known as {\it the uncertainty principle}. The inability to determine the coordinates of an arbitrary qubit is not an issue of insufficient measuring devices, but a consequence of the fundamental laws of quantum mechanics. According to these laws, an arbitrary qubit may be realized (up to a certain amount of precision, dependent of the accuracy of the equipment used), but it cannot be read. Qubits can be {\it measured}. Measurements of qubits are performed {\it in reference to a chosen orthonormal basis of} $\hil$ and the outcome of the measurement is random, either zero or one, as detailed below. As a result of the measurement, the qubit is transformed into one of the two qubits of the orthonormal basis chosen. The probability of obtaining each of the possible outcomes is the square of the absolute value of the corresponding coordinate of the qubit in the chosen basis. Explicitly, given $\theta,\varphi\in\mathbb{R}$, denote

\begin{equation}
\label{psi01}
\ket{\psi_0}=\Bigl( \begin{smallmatrix} \cos(\theta / 2) \\ e^{i\varphi}\sin(\theta / 2) \end{smallmatrix} \Bigr),\qquad \ket{\psi_1}=\Bigl( \begin{smallmatrix} \sin(\theta / 2) \\ -e^{i\varphi}\cos(\theta / 2) \end{smallmatrix} \Bigr),
\end{equation}

\noindent and denote by $B_{(\theta,\varphi)}$ the orthonormal basis $\{\ket{\psi_0},\ket{\psi_1}\}$ of $\hil$. Bloch sphere visualizations of $\ket{\psi_0}$ and $\ket{\psi_1}$ may be found in Figure 2.
\begin{center}
\includegraphics[scale=0.26]{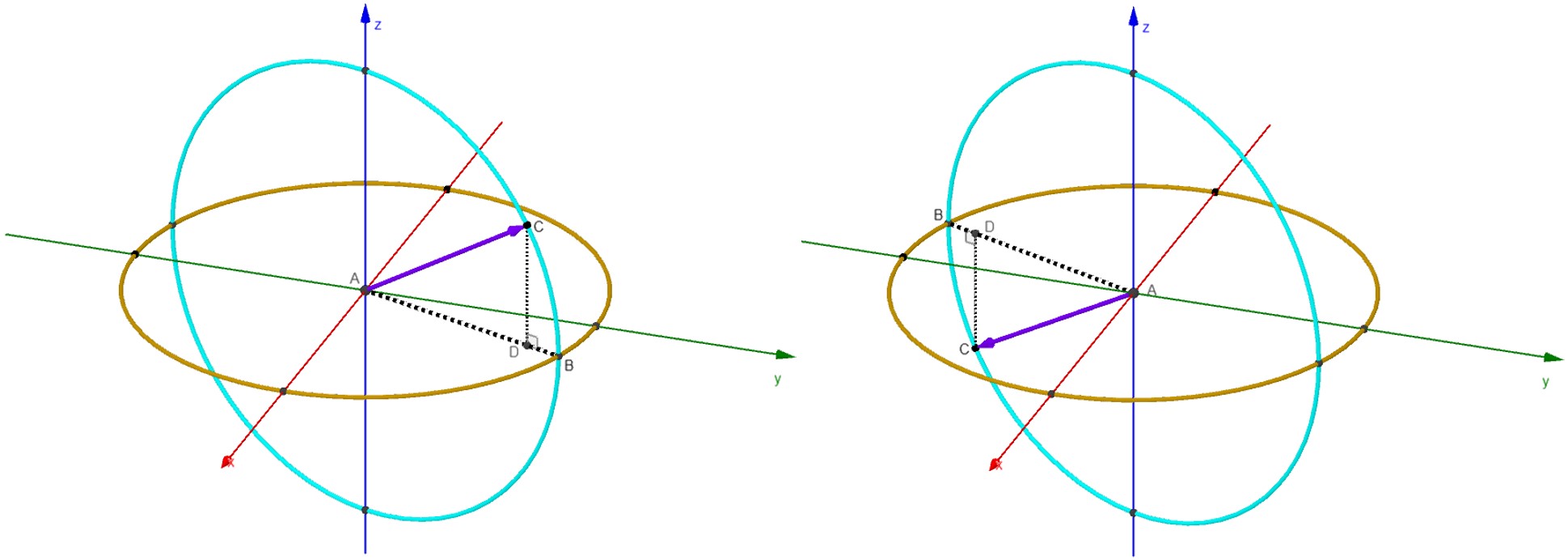}\\
\end{center}

\begin{center}
\begin{minipage}{0.8\linewidth}
\small \textit {Figure 2: Bloch sphere visualization of the orthonormal basis $B_{(\theta,\varphi)} = \{ \ket{\psi_0},\ket{\psi_1}\}$.} 
\end{minipage}
\end{center}
The fact that $\sin (\theta / 2) = \cos \frac{\pi -\theta}{2}$ and that $-e^{i\varphi}=e^{i(\pi-\varphi)}$ explains why the visualization of $\ket{\psi_1}$ is given by the negation of the visualization of $\ket{\psi_0}$. Visualizing an arbitrary qubit as a unit vector in $\mathbb{R}^3$, the measurement process may be thought of as choosing a point on the surface of the unit sphere in $\mathbb{R}^3$, piercing the solid (opaque) shell of the sphere, and ``looking into the sphere'' through that point (like looking through a door's key-hole). The measured qubit is immediately influenced by the observation and changes itself to point directly to the point of observation or to the exact opposite direction, in-deterministically. 
For a qubit $\ket{\psi}\in\hil$ and an orthonormal basis $B_{(\theta,\varphi)}$ of $\hil$, write $\ket{\psi}=\alpha \ket{\psi_0}+\beta \ket{\psi_1}$. When $\ket{\psi}$ is measured in reference to $B_{(\theta,\varphi)}$, there is a probability of $|\alpha|^2$ that $\ket{\psi}$ will transform into $\ket{\psi_0}$, yielding the outcome $0$, and a probability of $|\beta|^2$ that it will transform into $\ket{\psi_1}$, yielding the outcome $1$. We say that, when $\ket{\psi}$ is measured in reference to the basis $B_{(\theta,\varphi)}$, it {\it collapses} into one of the elements of that basis. Given $B_{(\theta,\varphi)}$, an orthonormal basis of $\hil$, any unit vector $\ket{\psi} = \alpha \ket{\psi_0} + \beta \ket{\psi_1}$ is {\it a superposition of $\ket{\psi_0}$ and $\ket{\psi_1}$}, and the elements of $B_{(\theta,\varphi)}$ are {\it pure states in reference to} $B_{(\theta,\varphi)}$. Since $B_{(\theta,\varphi)}$ is an orthonormal basis, $\alpha$ and $\beta$ are the inner products of $\ket{\psi}$ and the elements of $B_{(\theta,\varphi)}$. In general, if $B=\{\ket{v_1},\dots,\ket{v_n}\}$ is an orthonormal basis of an $n$-dimensional Hilbert space and $\ket{v}=\sum_{j=1}^n \alpha_j \ket{v_j}$, the inner product of $\ket{v_k}$ and $\ket{v}$, denoted by $\braket{v_k|v}$, is
\begin{equation}
\label{o-n-method}
\braket{v_k|v} = \Bigg \langle v_k \Bigg| \sum_{j=1}^n \alpha_j \ket{v_j}\Bigg \rangle=\sum_{j=1}^n \alpha_j \braket{v_k|v_j} = \alpha_k.
\end{equation}
Hence, $|\alpha|^2=|\braket{\psi_0|\psi}|^2$ and $|\beta|^2=|\braket{\psi_1|\psi}|^2$. This fact is used in this paper to compute the probabilities of the possible outcomes when measuring a given qubit (or a system of qubits) in reference to a given orthonormal basis. Measurements of systems of $l$ qubits are performed in reference to orthonormal bases of $\hil^{\otimes l}$, and result in a collapse of the system into one of the elements of that basis. The possible outcomes of such a measurement are the corresponding binary strings of length $l$, and the probability of obtaining each of the possible outcomes is the square of the absolute value of the corresponding coordinate of the system in the chosen basis. These may be computed using (\ref{o-n-method}). E.g., consider $l=2$, and let $B_{(\theta,\varphi)}=\{\ket{\psi_0},\ket{\psi_1}\}$ and $B_{(\theta',\varphi')}=\{\ket{\psi'_0},\ket{\psi'_1}\}$ two orthonormal bases of $\hil$. Tensor products of elements of these bases give the following orthonormal basis $\{\ket{\psi_0\psi'_0},\ket{\psi_0\psi'_1},\ket{\psi_1\psi'_0},\ket{\psi_1\psi'_1}\}$, denoted $B_{(\theta,\varphi)}\otimes B_{(\theta',\varphi')}$ of $\hil^{\otimes 2}$. Given a system of two qubits, measuring that system in reference to $B_{(\theta,\varphi)}\otimes B_{(\theta',\varphi')}$ is equivalent to measuring the first qubit in reference to $B_{(\theta,\varphi)}$ and the second qubit in reference to $B_{(\theta',\varphi')}$. \\

\section{The Random Basis Encryption Scheme}
\label{sec:4:2}
\subsection{The scheme.}
{\color{black}
We begin with some intuition. Our primary goal is to encrypt classical bits in a way that allows performing homomorphic operations over ciphertext. Our first step is taking a random pair of angles $(\theta,\varphi) \in[0,2\pi]^2$ and thinking of $\ket{\psi_0}=\begin{psmallmatrix} \cos(\theta / 2)\\ e^{i\varphi}\sin(\theta / 2) \end{psmallmatrix}$ as the encryption of the bit $0$ with $(\theta,\varphi)$ being the encryption key. Next, to allow homomorphic operations, we want that the encryption of the bit $1$ with the same key to be orthogonal to $\ket{\psi_0}$. Hence, we set $\ket{\psi_1}=\begin{psmallmatrix} \sin(\theta / 2) \\ -e^{i\varphi}\cos(\theta / 2)\end{psmallmatrix}$. Indeed, $\ket{\psi_0}$ and $\ket{\psi_1}$ are orthogonal and form an orthonormal basis of $\hil$. We denote this basis by $B_{(\theta,\varphi)}$. Next, to support fully compact, non-interactive, and IT-secure homomorphic $NOT$ operations, we want pairs $(\theta,\varphi)$ that satisfy $\ket{\psi_0}=NOT\ket{\psi_1}$ and $\ket{\psi_1}=NOT\ket{\psi_0}$ (up to a global phase factor). Solving these equations, we get $\varphi=\pm \pi /2$. This implies that for a pair of angles $(\theta,\varphi)\in[0,2\pi] \times \{\pm \frac{\pi}{2}\}$, the orthonormal basis $$B_{\bigl(\theta,\pm \tfrac{\pi}{2}\bigr)}=\Bigl\{\begin{psmallmatrix} \cos(\theta / 2)\\ \pm i\sin(\theta / 2) \end{psmallmatrix}, \begin{psmallmatrix} \sin(\theta / 2)\\ \mp i \cos(\theta / 2) \end{psmallmatrix}\Bigr\}$$ satisfies our first goal. It encrypts classical bits, and the quantum $NOT$ operation interchanges the encryptions in the same way that a classical (Boolean) negation gate interchanges the $0$ and $1$ bits. This observation, and the inability to determine the coordinates of an arbitrary qubit from a realization of it, lead us to the RBE QHE scheme of classical data. The RBE scheme allows a user to outsource the storage of confidential information to an untrusted server while enabling restricted homomorphic operations. First, we present the algorithms \verb|RBE.Gen|, \verb|RBE.Enc|, and \verb|RBE.Dec|. Next, we construct \verb|RBE.Eval|, and detail operations that may be homomorphically applied to the ciphertext in a fully compact and non-interactive way.

\begin{framed}
\begin{center}
    Random Basis Quantum Homomorphic Encryption
\end{center}

\noindent {\verb|RBE.Gen|:} 
\begin{itemize}
    \item Pick a pair of angles $(\theta,\varphi)$ from $[0,2\pi] \times \{\frac{\pi}{2},-\frac{\pi}{2}\}$ uniformly at random.
    \item Output $(\theta,\varphi)$.
\end{itemize}

\noindent{\verb|RBE.Enc|:} Given $b\in\m{M}$ and a $k=(\theta,\varphi)$:
\begin{itemize}
\item Denote by $K$ the $2$-by-$2$ complex matrix $\biggl( \begin{smallmatrix} \cos(\theta / 2) & \sin(\theta / 2)\\ e^{i\varphi}\sin(\theta / 2) & -e^{i\varphi}\cos(\theta / 2)\end{smallmatrix} \biggr)$.
\item Apply the quantum gate whose matrix representation is $K$ to a qubit $\ket{b}$ and obtain a new qubit.
\item Output the new qubit, denoted by $\ket{q}$.
\end{itemize}

\noindent {\verb|RBE.Dec|:} Given a qubit $\ket{\psi}$ and a pair of angles $(\theta,\varphi)$:
\begin{itemize}
\item Denote by $K^\dagger$ the conjugate transpose of $K$ ($K$ is as in {\verb|RBE.Enc|}). 
\item Apply the quantum gate whose matrix representation is $K^\dagger$ to $\ket{\psi}$ and obtain a new qubit.
\item Measure the obtained qubit in the computational basis and output the measurement's outcome.
\item Output the outcome of the measurement. 
\end{itemize}

\end{framed}
The columns of $K$ are the elements of $B_{(\theta,\varphi)}$. Multiplying the elements of the computational basis, $\{\ket{0},\ket{1}\}$, by $K$, we obtain the elements of $B_{(\theta,\varphi)}$. We refer to the encryption algorithm as taking the elements of the computational basis to the elements of the random basis $B_{(\theta,\varphi)}$. Since $K$ is a unitary transformation, $K^\dagger$ is its inverse, and hence, given ${(\theta,\varphi)}$, the decryption algorithm takes $B_{(\theta,\varphi)}$ elements to the computational basis elements. Of course, one may use the scheme to encrypt a string $x$ bit-by-bit and outsource $x$ to be stored and processed by a semi-trusted cloud. The scheme is perfectly correct. Indeed, assume that $\ket{q}$ is the encryption of $b\in\{0,1\}$ using ${(\theta,\varphi)}$. By \verb|RBE.Enc|, $\ket{q}=K\ket{b}$. In \verb|RBE.Dec|, $K^\dagger$ is applied to $\ket{q}$. One has $K^\dagger\ket{q}=K^\dagger K\ket{b}=\ket{b}$. Since $\ket{b}$ is a pure state, measuring it in reference to the computational basis, we get $b$ with probability~1. Below, we prove that the scheme is IT-secure. The key generation algorithm, \verb|RBE.Gen|, picks a key from an infinitely large domain. To implement it, one must make the domain finite. Remark 1 below discusses how $\m{K}$ may be made discrete and the security consequences of this procedure.}


\subsection{Security of the RBE scheme}
\label{sec:4:appB}

We now prove that the random basis encryption scheme is IT-secure. We do it in two different ways. First, as our scheme deals with encrypting and computing over classical data, we give a proof based on standard security definitions of classical schemes. Namely, we use a variant of a standard privacy definition from \cite{109}. The second proof follows a standard privacy definition from the quantum setting derived from \cite{122}.  

As described in Section~\RN{1}, an encryption scheme is composed of three algorithms, \verb|Gen|, \verb|Enc| and \verb|Dec|. $\m{M}$, $\m{K}$ and $\m{C}$ are the message space, key space and ciphertext space of the scheme, respectively. In our case, $\m{M}=\{0,1\}$ and $\m{K}=[0,2\pi]\times\{\pm \frac{\pi}{2}\}$. What is $\m{C}$? On the one hand, $\m{C}$ is the set of possible outputs of {\verb|Enc|}, implying that $\m{C}=\hil$. On the other hand, a ciphertext cannot indicate the encrypted information if it is not read. To read information from a qubit, one must measure that qubit. The output of such a measurement is an element of $\{0,1\}$, implying that $\m{C}=\{0,1\}$. The first (classical approach) proof uses the latter interpretation of $\m{C}$, and the second (quantum approach) proof uses the former. 

We begin with the classical approach. Assume that an adversary is holding an encryption $\ket{q}$ of $b$ generated using some key $(\theta,\varphi)\in\m{K}$. The adversary wishes to use $\ket{q}$ to find $b$, or to gain any information that will enable a better guess of $b$. The adversary is only able to measure $\ket{q}$ in reference to any orthonormal basis he chooses. If the measurement is performed in reference to any orthonormal basis other than $B_{(\theta,\varphi)}$, then each of the outcomes zero or one may be obtained with positive probability. We now rigorously prove that, no matter which orthonormal basis $B_{(\theta_0,\varphi_0)}$ is used by the adversary to measure $\ket{q}$, the probability of each of the outcomes zero or one is $\frac{1}{2}$, regardless of the actual value of $b$.

We now define the security criterion. Since \verb|Gen| is a probabilistic algorithm, given a message $m\in\m{M}$, the probability distribution over $\m{K}$ induces a probability distribution over $\m{C}$. An encryption scheme is {\it perfectly secure} if all messages $m\in\m{M}$ induce the same probability distribution over $\m{C}$. Formally (see \cite[Lemma 2.3]{109}):\\

\noindent {\bf Definition 1.} {\it An encryption scheme (\verb|Gen|, \verb|Enc|, \verb|Dec|) over a message space $\m{M}$ is perfectly secure if for every probability distribution over $\m{M}$, every $m_0,m_1\in\m{M}$, and every $c\in\m{C}$:
$$Pr[C = c | M = m_0] = Pr[C = c | M = m_1],$$
where $C$ and $M$ are the random variables denoting the value of the ciphertext and the message, respectively.}\\

By Definition 1, perfect security of the random basis encryption scheme follows from \\

\noindent {\bf Lemma 1.} { \it Let $(\theta_0,\varphi_0) \in [0,2\pi]^2$. One has

\begin{equation}
\label{securitycriteria}
Pr\bigl[{\mathbf M}\bigr(\ket{\psi_0},B_{(\theta_0,\varphi_0)}\bigl)=0\bigr] = Pr\bigl[{\mathbf M}\bigr(\ket{\psi_1},B_{(\theta_0,\varphi_0)}\bigl)=0\bigr],
\end{equation}
where 
\begin{itemize}
\item $B_{(\theta_0,\varphi_0)}$ is the orthonormal basis used by an adversary to measure an encryption of a bit,
\item $\ket{\psi_0}$ and $\ket{\psi_1}$ are as in (\ref{psi01}), and are encryptions of zero and one, obtained using our scheme, 
\item $\mathbf M \bigr(\ket{\psi},B_{(\theta_0,\varphi_0)}\bigl)$ is the random variable denoting the result obtained when measuring $\ket{\psi}$ in reference to $B_{(\theta_0,\varphi_0)}$,
\item the probability is over the choice of $(\theta,\varphi)$ from $[0,2\pi]^2$ and the inherent randomness of quantum measurements.
\end{itemize}}

\vspace{3mm}

\noindent {\bf Proof of Lemma 1.} We begin with computing the expression on the left-hand side $Pr\bigl[\mathbf M\bigr(\ket{\psi_0},B_{(\theta_0,\varphi_0)}\bigl)=0\bigr]$ of (\ref{securitycriteria}). That is, computing the probability of obtaining the outcome zero when measuring $\ket{\psi_0}$ in reference to $B_{(\theta_0,\varphi_0)}$ in terms of $\theta$ and $\varphi$. This probability is the square of the absolute value of the first coordinate of $\ket{\psi_0}$ in the orthonormal basis $B_{(\theta_0,\varphi_0)}$. Denote by $\ket{v_0}$ and $\ket{v_1}$ the elements of $B_{(\theta_0,\varphi_0)}$. As mentioned in (\ref{o-n-method}), the coordinates of $\ket{\psi_0}$ in $B_{(\theta_0,\varphi_0)}$ are given by appropriate inner products. Define $\alpha_0,\beta_0\in \C$ by $\ket{\psi_0}=\alpha_0\ket{v_0}+\beta_0\ket{v_1}$. One has
$$ \alpha_0=\braket{v_0|\psi_0} =\Braket{ \Bigl( \begin{smallmatrix} \cos(\theta_0 / 2) \\ e^{i\varphi_0} \sin(\theta_0 / 2) \end{smallmatrix} \Bigr) \Bigg| \Bigl( \begin{smallmatrix} \cos(\theta / 2) \\ e^{i\varphi} \sin(\theta / 2) \end{smallmatrix} \Bigr)} =\cos(\theta_0 / 2)\cos(\theta / 2)+e^{i(\varphi-\varphi_0)} \sin(\theta_0 / 2) \sin(\theta / 2).$$
Multiplying by $\alpha_0^*$, and using standard trigonometric identities, we obtain:
\begin{equation}
\label{t3}
|\alpha_0|^2 = \tfrac{1}{2}{\Bigl[}\cos ^2 \tfrac{\theta+\theta_0}{2}+\cos ^2 \tfrac{\theta-\theta_0}{2}+\sin \theta \sin \theta_0 \cos (\varphi-\varphi_0){\Bigr]}.
\end{equation}
Now, $\theta$ and $\varphi$ are chosen uniformly random from $[0,2\pi]\times\{\pm\frac{\pi}{2}\}$. The mean value of $|\alpha_0|^2$ over that domain may be computed in various ways. One may compute it using the formula $\overline{f}=\tfrac{1}{Vol(U)}\int_U f$, which yields $\tfrac{1}{2}$. By the law of total probability, the right-hand side of (\ref{securitycriteria}) is also $\tfrac{1}{2}$. All in all, we have 
\begin{center}
$Pr\bigl[\mathbf M\bigr(\ket{\psi_0},(\theta_0,\varphi_0)\bigl)=0\bigr] = Pr\bigl[\mathbf M\bigr(\ket{\psi_1},(\theta_0,\varphi_0)\bigl)=0\bigr]=\frac{1}{2}.$ $\square$
\end{center}
This concludes the classical proof. We have shown that, no matter which orthonormal basis is chosen by the adversary to measure $\ket{q}$, the outcome $0$ will be obtained with probability $\tfrac{1}{2}$, regardless of the actual value of $b$. By the laws of quantum mechanics, any operation other than measuring the qubit will yield less information regarding the plaintext. Since measuring the qubit gives no information at all, the scheme is perfectly secure. We now turn to the quantum approach, which interprets the ciphertext space as $\hil$. We use the density matrix representation of quantum states and base our claims on a security definition which follows the same line as Definition 3.1 from \cite{122} (modified for the continuous setting of our scheme). \\

\noindent {\bf Definition 2.} {\it Let $S\subseteq \hil$ be a set of qubits, $\mathcal{E}=\{U_i : i\in I\}$ be a set of unitary mappings on $\hil$, and $\rho_0$ be some density matrix. Uniformly at random applying an element of $\mathcal{E}$ to a given element $\ket{s}\in S$ perfectly hides $\ket{s}$ if and only if for all $\ket{s}\in S$ we have} 
$$ \int _I U_i \ket{s} \bra{s} U_i ^\dagger= \rho_0. $$

In our case, $S=\{\ket{0},\ket{1}\}$, and $\mathcal{E} = \biggl \{ \biggl( \begin{smallmatrix} \cos(\theta / 2) & \sin(\theta / 2)\\ e^{i\varphi}\sin(\theta / 2) & -e^{i\varphi}\cos(\theta / 2)\end{smallmatrix} \biggr) :  (\theta,\varphi) \in \m{K}\biggr\}$. To show that the random basis encryption scheme is perfectly secure, we need to show that 
\begin{equation}
\label{equation:int}
\int_{\m{K}} K_{\theta,\varphi} \ket{0} \bra{0} K_{\theta\varphi}^\dagger = \int_{\m{K}} K_{\theta\varphi} \ket{1} \bra{1} K_{\theta\varphi}^\dagger,
\end{equation}
where $K_{\theta\varphi}=\biggl( \begin{smallmatrix} \cos(\theta / 2) & \sin(\theta / 2)\\ e^{i\varphi}\sin(\theta / 2) & -e^{i\varphi}\cos(\theta / 2)\end{smallmatrix} \biggr)$. Standard computation shows that the left- and right-hand side of (\ref{equation:int}) are equal.  To conclude, the density matrix that an adversary sees after encryption is the same, regardless of the input. This shows that the random basis encryption scheme is perfectly secure. We note that, since the evaluation algorithm is non-interactive, the adversary gains no new information executing it, and hence the scheme is secure.\\

\noindent {\it Remark 1.} In the key generation algorithm of our random basis encryption scheme, the user is required to pick a uniformly random element $\theta$ from $[0,2\pi]$. Implementing random choices from a continuous domain might be technically challenging. However, the set of keys may be made discrete as follows. Let $N$ a positive integer, and $\mathcal{K}_N=\bigl \{ \tfrac{2\pi n}{N} : n \in \{1,2,\dots,N\} \bigr \}$. Instead of picking $\theta$ from $[0,2\pi]$, the user may uniformly at random pick $\theta$ from $\mathcal{K}_N$. How does that affect the security? In the classical security proof above, the mean value of the right hand side of (\ref{t3}) was computed by integrating over $[0,2\pi]$. Replacing $[0,2\pi]$ with $\mathcal{K}_N$, we compute the mean value of the right hand side of (\ref{t3}) by summing over all the possibilities for $\theta$ divided by $N$. Now, it is well known that for any real continuous function $f$,  
$$\int_{[0,2\pi]} f(x) dx = \lim_{N \to \infty} \sum_{n=1}^N \tfrac{2\pi}{N} f\bigl(\tfrac{2\pi n}{N}\bigr).$$
Hence, by taking large enough $N$, the mean value of the discrete version can be made arbitrarily close to $\frac{1}{2}$. In the quantum proof, by similar arguments, we can make the left- and right-hand sides of (\ref{equation:int}) arbitrarily close to each other by taking large enough $N$. To conclude, taking the discrete version of the key space, we make \verb|Gen| easier to implement in the cost of making the scheme statistically secure (rather than perfectly secure). Either way, the scheme is IT-secure. \\

\subsection{Homomorphic operations.}
We now explore the possibility of homomorphically applying quantum gates to the ciphertext by the untrusted quantum server. Obviously, any gate that commutes (up to a global phase factor) with the family of the encryption gates $K$, may be homomorphically applied to the encrypted data. Several unitary operations are typically used in quantum computing. We now investigate the consequences of applying some of these typically-used quantum gates to a random basis $B_{(\theta,\varphi)}$ encryption of classical data. As mentioned above, previous results show that IT-secure HE schemes (classical or quantum) cannot be efficient and support a universal set of gates at the same time \cite{105}. Hence, any  scheme that is both IT-secure and efficient (as our scheme is) cannot support a universal set of operations. However, we show that, our scheme can homomorphically support evaluation of several gates that will be useful for the applications presented below, in particular, WM-resilient QKD and securing entanglement. \\


\noindent{\it The ${NOT}$ gate.} The ${NOT}$ gate is the unitary transformation that interchanges the elements of the computational basis: $\ket{b} \rightarrow \ket{1-b}$. The matrix representation of ${NOT}$ in the computational basis is $X=\bigl(\begin{smallmatrix} 0 & 1 \\ 1 & 0 \end{smallmatrix}\bigr)$. What happens when one applies an $X$ gate to an element of a random basis $B_{(\theta,\varphi)}$? A simple calculation shows that, applying an $X$ gate to an element of $B_{(\theta,\varphi)}$ we get the other element of that basis, up to a global phase factor. Since $e^{i\varphi}=\pm i$, we have
$$X\ket{\psi_0} = \bigl(\begin{smallmatrix} 0 & 1 \\ 1 & 0 \end{smallmatrix}\bigr) \bigl( \begin{smallmatrix} \cos(\theta / 2) \\ \pm i \sin(\theta / 2) \end{smallmatrix} \bigr) = \bigl( \begin{smallmatrix} \pm i \sin(\theta / 2) \\ \cos(\theta / 2) \end{smallmatrix} \bigr)= \pm i \bigl( \begin{smallmatrix} \sin(\theta / 2) \\ \mp i \cos(\theta / 2) \end{smallmatrix} \bigr)= \pm i \ket{\psi_1}.$$
Similarly, $X\ket{\psi_1} =\mp \ket{\psi_0}$. To conclude, applying a $NOT$ gate to elements of $B_{(\theta,\varphi)}$ we get the same effect as when applying it to an element of the computational basis. Consequently, $X$ gates may be homomorphically applied to encrypted data.\\


\noindent{\it The Hadamard gate.} The Hadamard gate is the unitary transformation, whose matrix representation in the computational basis is $H=\tfrac{1}{\sqrt{2}}\bigl( \begin{smallmatrix} 1 & 1\\ 1 & -1\end{smallmatrix} \bigr)$. $H$ takes the elements of the computational basis to the elements of $B_{(\frac{\pi}{4},0)} = \Bigl\{ \frac{1}{\sqrt{2}}\bigl( \begin{smallmatrix} 1 \\ 1 \end{smallmatrix} \bigr), \frac{1}{\sqrt{2}}\bigl( \begin{smallmatrix} 1 \\ -1 \end{smallmatrix} \bigr)\Bigr\}$. The elements of $B_{(\frac{\pi}{4},0)}$ are often denoted by $\ket{+}$ and $\ket{-}$. One of the properties of the Hadamard gate is that, when measuring any of the elements of $B_{(\frac{\pi}{4},0)}$ in reference to the computational basis, the probabilities of obtaining zero or one are both~$\frac{1}{2}$. What are the probabilities of obtaining zero or one when measuring an element of $H\bigl[B_{(\theta,\varphi)}\bigr]$ in reference to $B_{(\theta,\varphi)}$? By Equation (\ref{o-n-method}), the probability of obtaining zero when measuring $H\ket{\psi_0}$ in reference to $B_{(\theta,\varphi)}$ is the square of the absolute value of the inner product of $H\ket{\psi_0}$ and $\ket{\psi_0}$. Since
\begin{equation}
H\ket{\psi_0}=\tfrac{1}{\sqrt{2}}\Bigl( \begin{smallmatrix} 1 & 1\\ 1 & -1\end{smallmatrix} \Bigr) \Bigl( \begin{smallmatrix} \cos(\theta / 2) \\ \pm i \sin(\theta / 2) \end{smallmatrix} \Bigr) =\tfrac{1}{\sqrt{2}}  \Bigl( \begin{smallmatrix} \cos(\theta / 2) \pm i \sin(\theta / 2) \\ \cos(\theta / 2) \mp i \sin(\theta / 2) \end{smallmatrix} \Bigr),
\end{equation}
the inner product is $\big\langle \psi_0\big| H \big|\psi_0 \big\rangle= \tfrac{\cos \theta}{\sqrt{2}}$. Hence, the probability of obtaining a zero outcome when measuring $H\ket{\psi_0}$ in reference to $B_{(\theta,\varphi)}$, is $\frac{\cos^2\theta}{2}$. Since the probabilities of the possible outcomes add up to one, when measuring $H\ket{\psi_0}$ in reference to $B_{(\theta,\varphi)}$ the outcome one is obtained with probability $\frac{1+\sin^2\theta}{2}$. Similar computations yield similar results for $\ket{\psi_1}$. Explicitly, when measuring $H\ket{\psi_1}$ in reference to $B_{(\theta,\varphi)}$, the probability of obtaining the outcome one is $\frac{\cos^2\theta}{2}$ and the probability of obtaining the outcome zero is $\frac{1+\sin^2\theta}{2}$. To conclude, applying a Hadamard gate to an element of a random basis, the probabilities of the elements of the basis in the superposition we get are in general not $\frac{1}{2}$ each.\\


These results are rather unfortunate since they indicate that the Hadamard gate does not create an equally weighted superposition when applied to an element of a random basis, and hence cannot be applied to the encrypted data homomorphically. Is there a quantum gate that takes elements of every $B_{(\theta,\varphi)}$ basis to an equally weighted superposition of the elements of that basis? The answer is `Yes'. Indeed, the following quantum gate satisfies this requirement using an ancillary $\ket{0}$ qubit:  

$$D=\tfrac{1}{\sqrt{2}}  \begin{psmallmatrix}  1  &  0 &  1 &  0 \\ 0  &  1 & 0 & 1 \\ 0  &  1 & 0 & -1 \\ 1  &  0 & -1 & 0 \end{psmallmatrix} .$$

\noindent $D$ is the matrix representation (in the computational basis) of the quantum gate used in \cite{117} to create {\it Bell states}. This gate is the two-qubit quantum circuit established by first applying a Hadamard gate to the first qubit, and then a $CNOT$ gate to that system of two qubits, where the first qubit is the control qubit and the second is the target qubit. That circuit is illustrated in Figure~3. 

\begin{center}
\includegraphics[scale=0.27]{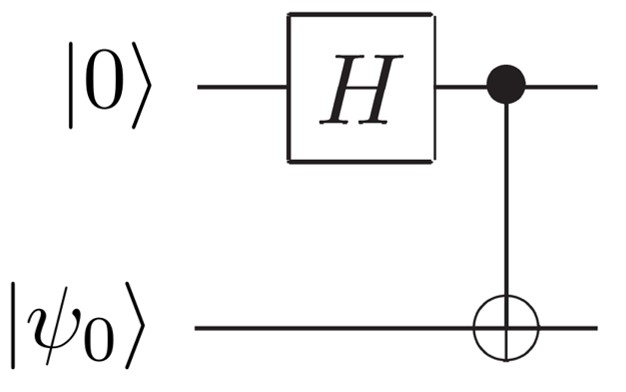}\\
%
\begin{minipage}{0.3\linewidth}
\small \textit {Figure 3: Random Based $D$ gate.} 
\end{minipage}
\end{center}

We now prove that, applying a $D$ gate to a tensor product of $\ket{0}$ and an element of a random basis, measuring the second qubit in reference to that same random basis, the probabilities of obtaining the outcomes zero and one are both $\tfrac{1}{2}$. Explicitly, let $\ket{\psi_b}$ an element of a random basis, $B_{(\theta,\varphi)}$, where $\varphi = $ and $\theta\in[0,2\pi]$. We have

\noindent {\bf Lemma 2.} {\it $D$ is a quantum gate which takes tensor products of the form $\ket{0\psi_b}$  to a system of two qubits, such that, measuring that system in reference to $\{\ket{0},\ket{1}\}\otimes B_{(\theta,\varphi)}$, the probability of each of the outcomes zero and one for the second qubit is $\tfrac{1}{2}$.}

\noindent {\bf Proof of Lemma 2.}
\noindent {\it Proof.} Let $\theta\in[0,2\pi]$ and $\varphi=\pm i$. One has:

\begin{equation}
\begin{aligned}
\label{D1}
\ket{0\psi_0}&= \bigl(\begin{smallmatrix} 1 \\ 0 \end{smallmatrix} \bigr) \otimes \biggl( \begin{smallmatrix} \cos(\theta / 2) \\ \pm i \sin(\theta / 2) \end{smallmatrix} \biggr) =  \begin{psmallmatrix} \cos(\theta / 2) \\ \pm i \sin(\theta / 2) \\ 0 \\ 0 \end{psmallmatrix} , \\
D \ket{0\psi_0}&=\tfrac{1}{\sqrt{2}} \begin{psmallmatrix} 1 & 0 & 1 & 0 \\ 0 & 1 & 0 & 1 \\ 0 & 1 & 0 & -1 \\ 1 & 0 & -1 & 0 \end{psmallmatrix} \begin{psmallmatrix} \cos(\theta / 2) \\ \pm i \sin(\theta / 2) \\ 0 \\ 0 \end{psmallmatrix} = 
\tfrac{1}{\sqrt{2}} \begin{psmallmatrix}  \cos(\theta / 2) \\ \pm i \sin(\theta / 2) \\ \pm i \sin(\theta / 2)  \\ \cos(\theta / 2) \end{psmallmatrix}.
\end{aligned}
\end{equation}

\noindent The probabilities of obtaining each of the possible outcomes, when measuring $D \ket{0\psi_0}$ in reference to $\{\ket{0},\ket{1}\}\otimes B_{(\theta,\varphi)}$, are the squares of the absolute values of the coordinates of $D \ket{0\psi_0}$ in that basis. The elements of $\{\ket{0},\ket{1}\}\otimes B_{(\theta,\varphi)}$ are $\ket{0\psi_0},\ket{0\psi_1},\ket{1\psi_0}$ and $\ket{1\psi_1}$.
The first, $\ket{0\psi_0}$, has been computed in (\ref{D1}). Now,

\begin{equation}
\label{D2}
\ket{1\psi_1} =\bigl(\begin{smallmatrix} 0 \\ 1 \end{smallmatrix} \bigr) \otimes \biggl( \begin{smallmatrix} \sin(\theta / 2) \\ \mp i \cos(\theta / 2) \end{smallmatrix} \biggr)=
\begin{psmallmatrix} 0 \\ 0 \\ \sin(\theta / 2) \\ \mp i \cos(\theta / 2)  \end{psmallmatrix}.
\end{equation}

\noindent By (\ref{D1}) and (\ref{D2}), 
$$\frac{\ket{0\psi_0}\pm i \ket{1\psi_1}}{\sqrt{2}}= \tfrac{1}{\sqrt{2}} \begin{psmallmatrix}  \cos(\theta / 2) \\ \pm i \sin(\theta / 2) \\ \pm i \sin(\theta / 2)  \\ \cos(\theta / 2) \end{psmallmatrix}=D \ket{0\psi_0}.$$

\noindent This shows that the coordinates of $D \ket{0\psi_0}$ in $\{\ket{0},\ket{1}\}\otimes B_{(\theta,\varphi)}$ are $\tfrac{1}{\sqrt{2}},0,0$ and $\tfrac{\pm i}{\sqrt{2}}$. Taking the squares of the absolute values of these coordinates one sees that, measuring in reference to $\{\ket{0},\ket{1}\}\otimes B_{(\theta,\varphi)}$, the outcome $00$ is obtained with probability $\tfrac{1}{2}$, as so is $11$. The probabilities of obtaining the different outcomes when measuring $D\ket{0\psi_1}$ in reference to $\{\ket{0},\ket{1}\}\otimes B_{(\theta,\varphi)}$ may be found by substituting $\theta = \pi-\theta'$ and $\varphi=-\varphi'$. That substitution yields $D\ket{0\psi_1}=\frac{\ket{0\psi_1}\mp i \ket{1\psi_0}}{\sqrt{2}}$. Taking the squares of the absolute values, we obtain the desired probabilities. $\square$\\

To conclude, the $D$ gate may be homomorphically applied to the elements of a random basis, using an ancillary $\ket{0}$ qubit, resulting in the same effect as when applying a Hadamard gate to the elements of the computational basis -- creating a superposition of the elements of that basis with equal probabilities. We note that the ancillary qubit may be generated by the server with no interference of or interaction with the user. We stress that, the $D$ gate presented above is not intended to emulate a Hadamard gate completely. It only takes elements of orthonormal bases to equally weighted superpositions of the states. However, this attribute is found to be sufficient for the applications presented below.\\


\noindent{\it The $CNOT$ gate.} The $CNOT$ gate is a two-qubit gate, whose matrix representation in the computational basis of $\hil^{\otimes 2}$ is $$\biggl( \begin{smallmatrix} 1 & 0 & 0 & 0 \\ 0 & 1 & 0 & 0 \\ 0 & 0 & 0 & 1 \\ 0 & 0 & 1 & 0 \end{smallmatrix} \biggr).$$ 

\noindent Tensor products of the elements of the computational basis $\{\ket{0},\ket{1}\}$ of $\hil$, give the computational basis $\{\ket{00}, \ket{01}, \ket{10}, \ket{11} \}$ of $\hil^{\otimes2}$. 
Applying the $CNOT$ gate to the elements of the latter basis, we leave $\ket{00}$ and $\ket{01}$ unchanged, and interchange $\ket{10}$ and $\ket{11}$. In other words, if the first qubit is $\ket{0}$, then the second qubit is left unchanged, and if the first qubit is $\ket{1}$, then a $NOT$ gate is applied to the second qubit. For this reason, this gate is called {\it the controlled-$NOT$ gate}. The first qubit is {\it the control qubit} and the second is {\it the target qubit}. 

What happens if one applies a $CNOT$ gate to the elements of a random basis of $\hil^{\otimes2}$? Namely, let $B_{(\theta,\varphi)}=\{\ket{\psi_0},\ket{\psi_1}\}$ and $B_{(\theta',\varphi')}=\{\ket{\psi'_0},\ket{\psi'_1}\}$ two orthonormal bases of $H$. Tensor products of the elements of $B_{(\theta,\varphi)}$ and $B_{(\theta',\varphi')}$ give the following orthonormal basis of $\hil^{\otimes2}$: $$\{\ket{\psi_0\psi'_0},\ket{\psi_0\psi'_1},\ket{\psi_1\psi'_0},\ket{\psi_1\psi'_1}\}.$$ 

Is the {\it control-target structure} kept when applying $CNOT$ to the elements of that basis, leaving $\ket{\psi_0\psi'_0}$ and $\ket{\psi_0\psi'_1}$ unchanged, and interchanging $\ket{\psi_1\psi'_0}$ and $\ket{\psi_1\psi'_1}$? The answer turns out to be negative. Applying a $CNOT$ gate to these elements, we take each of them to a superposition of the others. \\

Can we find a quantum gate (using ancillary qubits, perhaps) that keeps the control-target structure when applied to the elements of a random basis of $\hil^{\otimes2}$? The answer is negative. Indeed, assume by contradiction that $P$ is such a gate. $P$ must leave $\ket{\psi_0 \psi_0}$ unchanged and take $\ket{\psi_1 \psi_1}$ to $\ket{\psi_1 \psi_0}$, regardless of $\theta$ and $\varphi$. Let $\theta'=\pi - \theta$ and $\varphi'=\pi-\varphi$. Switching between $\ket{\psi_0}$ and $\ket{\psi_1}$ and examining $P$'s operation on $\ket{\psi_0 \psi_0}$ and $\ket{\psi_1 \psi_1}$ implies a contradiction. Indeed, w.l.o.g., consider the following two cases. First, if $\theta =0$ and $\varphi=\pi$, we have $\ket{\psi_0}=\ket{0}$ and $\ket{\psi_1}=\ket{1}$. Second, if $\theta =\pi$ and $\varphi=0$, we have $\ket{\psi_0}=\ket{1}$ and $\ket{\psi_1}=\ket{0}$. In the first case, $P\ket{\psi_0 \psi_0}=P\ket{00}$ and $P\ket{\psi_1 \psi_1}=P\ket{11}$, implying that $\ket{00}$ is unchanged by $P$ and $\ket{11}$ is taken to $\ket{10}$. On the other hand, in the second case, $P\ket{\psi_0 \psi_0}=P\ket{11}$ and $P\ket{\psi_1 \psi_1}=P\ket{00}$, implying that $\ket{11}$ is unchanged and $\ket{00}$ is taken to $\ket{01}$. By the first case, $\ket{00}$ is unchanged by $P$, but by the second case, $P$ takes it to $\ket{01}$ -- a contradiction! This shows that such a $P$ cannot exist. 

Nevertheless, by applying a $CNOT$ gate to the elements of a {\it partially-random} basis $\{\ket{0},\ket{1}\}\otimes B_{(\theta,\varphi)}$ of $\hil^{\otimes2}$ we do keep the target-control structure. The elements of such a basis are 
$$\ket{0\psi_0} = \begin{psmallmatrix} \cos (\theta / 2) \\ \pm i\sin (\theta / 2) \\ 0 \\ 0  \end{psmallmatrix}, \ket{0\psi_1} = \begin{psmallmatrix} \sin (\theta / 2) \\ \mp i\cos (\theta / 2) \\ 0 \\ 0  \end{psmallmatrix} , \ket{1\psi_0} = \begin{psmallmatrix}  0 \\ 0 \\ \cos (\theta / 2) \\ \pm i\sin (\theta / 2)   \end{psmallmatrix} ,\ket{1\psi_1} = \begin{psmallmatrix} 0\\0\\ \sin (\theta / 2) \\ \mp i\cos (\theta / 2)   \end{psmallmatrix} .$$ Applying a $CNOT$ gate to these elements, we leave $\ket{0\psi_b}$ unchanged and interchange $\ket{1\psi_b}$ and $\ket{1\psi_{1-b}}$, up to a global phase factor. In fact,

\begin{equation}
\label{cnot1}
CNOT \ket{1\psi_0} = 
\biggl( \begin{smallmatrix} 1 & 0 & 0 & 0 \\ 0 & 1 & 0 & 0 \\ 0 & 0 & 0 & 1 \\ 0 & 0 & 1 & 0 \end{smallmatrix} \biggr)
\Biggl(\begin{smallmatrix}  0 \\ 0 \\ \cos (\theta / 2) \\ \pm i\sin (\theta / 2)   \end{smallmatrix} \Biggr) = 
\Biggl(\begin{smallmatrix}  0 \\ 0 \\ \pm i\sin (\theta / 2) \\ \cos (\theta / 2)  \end{smallmatrix} \Biggr) = 
\pm i \Biggl(\begin{smallmatrix} 0\\0\\ \sin (\theta / 2) \\ \mp i\cos (\theta / 2)   \end{smallmatrix} \Biggr) = 
\pm i \ket{1\psi_1}.
\end{equation}
A similar computation shows that $CNOT \ket{1\psi_1} = \mp i \ket{1\psi_0}$. Since the last two entries of $\ket{0\psi_b}$ are zero, applying a $CNOT$ gate we leave them unchanged. To conclude, $CNOT$ gates may be homomorphically applied to systems of two qubits when the control qubit is an element of the computational basis and the target qubit is an element of $B_{(\theta,\varphi)}$. This property of our scheme is found to be useful for applications presented below.\\

\noindent{\it $C^nNOT$ gates.} For a positive integer $n$, the $C^nNOT$ gate is an $n+1$ qubit gate, whose matrix representation in the computational basis of $\hil^{\otimes(n+1)}$ is the matrix obtained from the identity matrix of order $2^{n+1}$ by replacing its bottom right block $\begin{psmallmatrix} 1 & 0 \\ 0 & 1 \end{psmallmatrix}$ with the block $\begin{psmallmatrix} 0 & 1 \\ 1 & 0 \end{psmallmatrix}$. Namely, the $NOT$ and $CNOT$ gates discussed above are the special cases $n=0$ and $n=1$, respectively, of $C^nNOT$. Similarly to (\ref{cnot1}), one may readily verify that, given a random basis $B_{(\theta,\varphi)}$,

\begin{equation}
\label{cnnot}
C^nNOT\ket{b_1b_2\dots b_n\psi_b} =
\begin{cases} 
\ket{b_1b_2\dots b_n\psi_{1-b}}, &\mbox{} \qquad \prod_{i=1}^n b_i=1,\\
\ket{b_1b_2\dots b_n\psi_b}, & \mbox{\qquad otherwise.} 
\end{cases}
\end{equation}
Hence, $C^nNOT$ gates may be homomorphically applied to systems of qubits when the control qubits are elements of the computational basis and the target qubit is an element of $B_{(\theta,\varphi)}$.

To conclude, we have shown that our RBE scheme supports homomorphic $NOT$ operations, and $CNOT$ gates, where the control qubits are set in clear. It also supports the $D$ gate -- a quantum gate that uses an ancillary qubit to take the elements of an orthonormal basis to an equally weighted superposition of the states. Below, we show the usefulness of these attributes for several applications.

\section{Securing entanglement}
\label{sec:4:3}

Entanglement is an essential resource in quantum computation. Once generated, it should be guaranteed that only the rightful owners of it would be able to use it. In this section, we present a method for securing that important resource in an IT-secure way, using our scheme. One example of a setting in which entanglement is used as a core element is {\it Quantum Pseudo-Telepathy} games. This concept was introduced in \cite{115} and refers to the use of entanglement to eliminate the need for communication in specific multiparty tasks. Comprehensive coverage of the subject may be found in \cite{116}. An example of such a task is the {\it Mermin-Peres magic square game} \cite{118}. In this game, two parties, Alice and Bob, are presented with a 3$\times$3 table. Each of them is required to fill in a part of the table, as follows. Alice is given a number $i$, $1 \leq i \leq 3$, and needs to put either 0 or 1 at each entry of the $i$-th row, in such a way that the sum of the three entries will be even. Similarly, Bob is given a $j$, $1 \leq j \leq 3$, and needs to fill in the $j$-th column with the constraint that the sum be odd. The numbers $i$ and $j$ are the inputs of the parties. Alice and Bob win the game if they place the same number at the intersection of the row and the column that they fill. The parties do not know $i$ and $j$ ahead of the game, and they cannot communicate after being given these values. They are allowed to communicate before the game begins and discuss game strategies, or share any information they desire. It was shown in \cite{116} that there is no classical algorithm that lets Alice and Bob win the game with probability greater than $\frac{8}{9}$, whereas there exists a quantum algorithm that lets them win the game with probability 1. This quantum algorithm is based on having each of the parties hold two qubits out of an entangled system of four qubits. The system of four qubits used in \cite{118} for that purpose is
$$\ket{\Psi}= \tfrac{1}{2}\ket{0011}-\tfrac{1}{2}\ket{0110}-\tfrac{1}{2}\ket{1001}+\tfrac{1}{2}\ket{1100}.$$

Assume that Alice and Bob are two parties that wish to engage in the magic square game. Alternatively, Alice and Bob are two scientists working in distant labs and wish to complete a joint task that requires entanglement. First, we consider the case in which Alice and Bob can get together and jointly generate entangled qubits, or purchase them from a trusted provider. Alice and Bob, having obtained a large number of entangled qubits, store these qubits in their laboratories to use them when the task requires it. Alice and Bob are worried that at the end of the day, when Alice and Bob are no longer at their labs, other people, say, Eve and Mallory will break into their labs. Eve will steal half of each entangled system from Alice's lab, and Mallory will steal the corresponding half from Bob's lab. Eve and Mallory will use the stolen entangled pairs for their own needs. In light of this concern, Alice and Bob are looking for a way to secure their entangled particles to ensure that no one else can use them. Like a smartphones password lock that will not let anyone use the smartphone without knowing the password.\\

\noindent {\bf First approach based on QOTP.} One may suggest that, before leaving their laboratories, Alice and Bob use QOTP to (independently) encrypt each half of each entangled pair. How will it work? For example, assume that Alice and Bob hold two halves of an EPR pair,

$$\ket{\Phi^+} = \tfrac{1}{\sqrt{2}} \bigl( \ket{0}_A  \ket{0}_B +  \ket{1}_A \ket{1}_B \bigr).$$

The subscripts $A$ and $B$ indicate the parts of the system held by each party. Alice picks QOTP keys $(a_1,a_2)$ uniformly at random from $\{0,1\}^2$, and Bob similarly picks $(b_1,b_2)$. At the end of the day, to secure the entangled pair, Alice applies $X^{a_1}Z^{a_2}$ to her half, and Bob applies $X^{b_1}Z^{b_2}$ to his part. Doing so, they obtain a new state: 

$$X^{a_1}Z^{a_2} \otimes X^{b_1}Z^{b_2} \ket{\Phi^+}= \tfrac{1}{\sqrt{2}} \Bigl( \bigl( X^{a_1}Z^{a_2} \ket{0}_A\bigr)  \bigl(X^{b_1}Z^{b_2} \ket{0}_B) \bigr) + 
\bigl( X^{a_1}Z^{a_2} \ket{1}_A\bigr)  \bigl(X^{b_1}Z^{b_2} \ket{1}_B) \bigr) \Bigr).$$

Since the encryption keys were picked uniformly at random and independently of each other, the density matrix of the new state is equal to the identity (up to a constant). So it seems like this procedure secures the entangled system in the sense that, without knowing the encryption keys, the encrypted system contains zero amount of entanglement. This claim can be phrased using conventional measures of entanglement like entanglement distillation and entanglement dilution. However, if Eve and Mallory steal a large amount of QOTP-encrypted EPR pairs from Alice and Bob, then they could guess the encryption keys for each pair, and their guess is expected to be correct $\tfrac{1}{16}$ of the times (on average). 

We want to refine this point. Eve and Mallory cannot produce two halves of an entangled system by using local operations and classical communication (LOCC) alone. Stealing QOTP-encrypted EPR pairs from Alice and Bob, they can recover the original entangled system with a non-negligible probability using LOCC alone. Then, the recovered systems can be used by Eve and Mallory for their purpose. We conclude that QOTP encryption of EPR pairs reduces the value of a stolen pair to $\tfrac{1}{16}$ of its original value. In such a situation, it still pays for Eve and Mallory to steal EPR pairs, as $6.25\%$ of them are expected to be usable. We assume that Eve and Mallory know that the entangled state encrypted by Alice and Bob is $\ket{\Phi^+} = \tfrac{\ket{00}+\ket{11}}{\sqrt{2}}.$ 

At this point we note that, When Eve and Mallory attempt to decrypt the pairs by randomly guessing the keys, they cannot tell which of the pairs are decrypted correctly. Nevertheless, one may suggest a scenario where, after using the randomly-decrypted pair, it is possible to tell if the guess was right and gain value from the outcome that could not have been gained from LOCC alone.   \\

\noindent {\bf Second approach based on RBE.} An alternative (and arguably better) way of securing entangled systems comes from our RBE scheme. We suggest that Alice and Bob use our RBE scheme to encrypt each half of an EPR pair using independent random keys $\theta_a$ and $\theta_b$. This way, if Eve and Mallory steal the encrypted qubits and try to decrypt them by guessing the keys, their guess is expected to be correct zero percent of the time. This may make stolen EPR pairs unusable, and in such a situation, the theft of EPR pairs becomes unprofitable. 

\noindent {\bf Secure transmission of self-generated entangled systems.} What happens if Alice and Bob are far apart and cannot get together to generate (or purchase) an entangled system? Being far apart, they may ask a third party, Charlie, to generate such an entangled system and transmit half of it to each of them. In that case, two concerns may arise. First, Charlie might be untrustworthy. Second, Eve and Mallory might intercept Charlie's transmission and use the entangled qubits sent by Charlie for their purposes (see Figure 4).

To overcome the possibility that Charlie is untrustworthy, Alice and Bob may decide that one of them, say, Alice will generate the entangled system and transmit half of it to Bob. This does not solve the second concern. A single adversary, Eve, may still intercept the transmission and use the half sent to Bob to engage in the task with Alice (instead of Bob, see Figure 5).

\begin{center}
\includegraphics[scale=0.3]{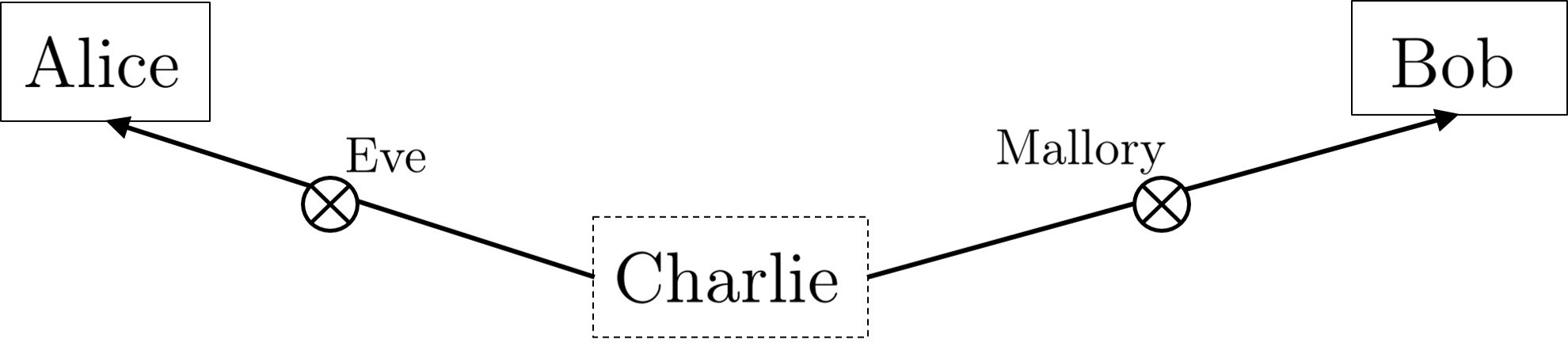}\\
\begin{minipage}{0.45\linewidth}
\small \textit {Figure 4: Adversarial attack by two adversaries.} 
\end{minipage}
\end{center}

To solve both concerns, Alice and Bob can securely generate and share an EPR pair using our random basis encryption scheme, as follows. 
\begin{itemize}
\item Alice generates an EPR pair and encrypts each half independently using our RBE scheme.  
\item Alice keeps the first half to herself and transmits the second half to Bob. 
\item Alice and Bob communicate through a secure communication channel (possibly, using our QKD scheme presented below) and Alice shares with Bob the key she used to encrypt his half. 
\item When they need to use the entangled system, Alice and Bob decrypt the qubits they hold and obtain a proper entangled system. 
\end{itemize}

This way, even if Eve intercepts the transmitted qubit, she can not use it to engage in the task instead of Alice without knowing the encryption key. 

\begin{center}
\includegraphics[scale=0.3]{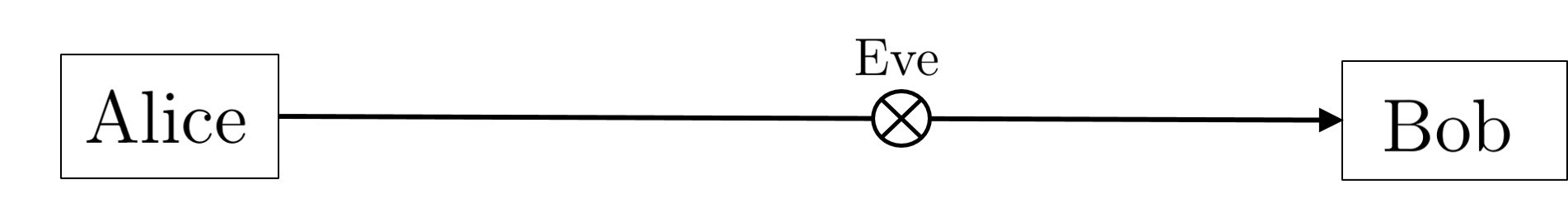}\\
\begin{minipage}{0.45\linewidth}
\small \textit {Figure 5: Adversarial attack by a single adversary.} 
\end{minipage}
\end{center}

\section{The RBE-based CNOT Quantum Key Distribution scheme} 
\label{sec:4:4}

Quantum key distribution (QKD), first suggested by Bennett and Brassard in 1984, is one of the most celebrated results in quantum computing. The discovery that quantum mechanics enables two distant parties to agree on a joint encryption key while relying on no computational assumptions is one of the most significant breakthroughs in the research on secure communications. However, the BB84 protocol, and most of the QKD schemes that followed it, do not prevent an eavesdropper from gaining any information on the key. Instead, these schemes are designed to enable Alice and Bob to {\it detect} eavesdropping attempts with high probability. This is done based on a fundamental postulate of quantum mechanics -- ``information gain is possible only at the cost of disturbing the state''. After invoking the quantum part of the QKD scheme, Alice and Bob invoke classical {\it privacy amplification} (PA) and {\it data reconciliation} (DR) procedures. These procedures are required to reduce the amount of information held by a possibly undetected eavesdropper, and to correct possible errors in the key caused by the eavesdropping (or by implementation errors). However, these procedures reduce the bandwidth and have time, communication, and computational costs. Similarly to the securing entanglement scenario, it would be very helpful if there was a way of reducing the ability of an eavesdropper to gain information in the first place, thereby impairing the motivation to attack the transmission and avoiding the expensive PA and DR procedures. \\

In this section, we review two QKD protocols, namely, the BB84 protocol and the QKD scheme suggested by Deng and Long in \cite{137}, and suggest a new type of attack against these schemes. Our attack is based on {\it weak measurements} (WM), and it enables the attacker, Eve, to control the probability in which Alice and Bob detect her. Our WM attack allows Eve a tradeoff between the probability of being caught and the amount of information that she can gain in her attack. Then, we introduce our RBE-based CNOT QKD scheme. Our QKD scheme, being resilient against such WM attacks, takes a step towards significantly impairing the motivation of a possible adversary.   \\


\noindent{\bf Reviewing the BB84 scheme.} We now briefly review the BB84 QKD scheme (described in detail also in \cite{103}). Alice picks two uniformly random bits $a$ and $b$ and generates the qubit $H^aX^b\ket{0}$, where $b$ is the bit to be transmitted, and $a$ determines the basis used to encode the bit. If $a=0$, the bit $b$ is encoded in the computational basis, and if $a=1$, the bit $b$ is encoded in the Hadamard basis. Alice transmits the qubit to Bob, who picks a uniformly random bit $c$, applies a $c$-conditioned Hadamard to the qubit, and measures the qubit in the computational basis. The bit $c$ is Bob's guess regarding the value of $a$. If Bob's guess about the basis is correct (i.e., $a=c$), Alice and Bob use the corresponding $b$ to either check for eavesdropping or generate the key. If Bob's guess is wrong, the corresponding $b$ is ignored. An illustration of the process is displayed in Figure 6. 

\begin{center}
\includegraphics[scale=0.65]{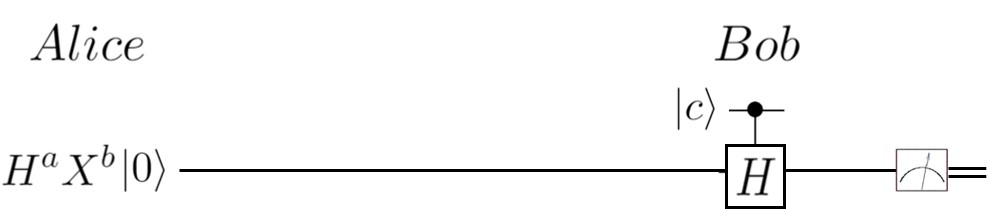}\\
\begin{minipage}{0.45\linewidth}
\small \textit {Figure 6: The BB84 QKD protocol.} 
\end{minipage}
\end{center}

Alice and Bob then announce $a$ and $c$. If Bob's guess is correct (i.e., $a=c$), which is expected with probability $0.5$, and there was no adversarial interference on the transmission, then the outcome of Bob's measurement is guaranteed to be $b$ (assuming an error-free quantum channel). Alice and Bob repeat this process for $\approx 4n$ qubits. On average, it is expected that $a=c$ for $\approx 2n$ of them. Then, to detect possible eavesdropping attempts, Alice and Bob publicly compare $n$ outcomes of Bob's measurements (randomly chosen from the $2n$ qubits for which $a=c$) with the corresponding $b$'s. If the error rate is too high, Alice and Bob abort. If not, they can bound the amount of information held by an eavesdropper, and then invoke PA and DR procedures to obtain a joint secure key. \\

\noindent{\bf Remark 2.} We observe a subtle issue that was apparently overlooked in previous works that studied the BB84 scheme. Consequently, we suggest an improvement to the BB84 scheme. Instead of Bob {\it guessing} the basis Alice used (by randomly choosing $c\in \{0,1\}$ and hoping that $c=a$), we suggest that Bob will notify Alice that he received the qubit, and then Alice will reveal the basis she used (i.e., reveal $a$).Next, Bob will set $c=a$, and continue the scheme from this point as specified above. Why is this variation results in a scheme which is still secure? Alice's basis (i.e., $a$) is announced only {\it after} the qubit has already arrived at Bob's safe hands, and hence Eve cannot use $a$ to gain any valuable information. Why is this variation helpful? Announcing the correct basis by Alice (i.e., revealing $a$) makes Bob's measurements performed according to the right basis $100\%$ of the time (instead of 50\%), which results in doubling the key generation rate. In other words, in the original BB84 scheme, Bob guesses the value of $a$ uniformly at random, and hence the corresponding $b$ is useful only 50\% of the time. We observe that, instead of guessing the basis, Bob may be informed by Alice on the basis after receiving the qubit, thereby doubling the efficiency. \\

\noindent{\bf A weak-measurement attack on the BB84 scheme.} {\color{black} Below we suggest an attack on the BB84 QKD protocol based on weak measurements. As mentioned in the Introduction, weak measurements are performed in two steps. The first step is to interact the target qubit with an ancilla. The second step includes (strongly) measuring the ancilla. The outcome of the weak measurement of the target qubit is the outcome of the (strong) measurement of the ancilla. Below, we give an example of such a procedure. For $\varepsilon >0$, denote 

$$\W=\sqrt{\varepsilon}\cdot i \cdot CNOT + \sqrt{1-\varepsilon} \cdot I\otimes I,$$ 

where $I$ is the identity over a single qubit. One readily verifies that $\W$ is unitary. An adversary can use $\W$ to extract information about bits transmitted in the BB84 protocol while leaving minor indications of the attack. We begin with some intuition. The ``no cloning theorem'' states that, in general, qubits cannot be cloned. Nonetheless, computational basis qubits {\it can} be cloned. Indeed, applying a $CNOT$ gate to a computational-basis qubit and an ancilla ($\ket{0}$) copies the qubit to the ancilla without disturbing the qubit. But if the qubit is not in the computational basis, the $CNOT$ gate {\it does} disturb it (and the qubit cannot be copied). $\W$ is a superposition of the identity and $CNOT$ gates. As $\varepsilon$ gets closer to $0$, $\W$ gets closer to the identity. Our weak measurement (WM) attack goes as follows. Given a qubit $\ket{\psi}$ on the computational or Hadamard basis, we suggest applying $\W$ to $\ket{\psi}$ and an ancilla and then measuring the ancillary qubit. If $\ket{\psi}$ is in the computational basis, the weak measurement yields information about $\ket{\psi}$, and if $\ket{\psi}$ is in the Hadamard basis, we slightly disturb the qubit.\\}

The WM-attack is formalized via the {\it key bit guessing game}. This game attempts to encapsulate the essence of a QKD scheme being IT-secure against eavesdropping attempts, and to measure the advantaged that can be gained from different attack strategies against QKD schemes. It is frequently the case that, as part of a QKD protocol, Alice and Bob use DR protocols (which are, essentially, error-correcting codes) and PA schemes (which are, essentially, cryptographic hash functions). However, our game measures the amount of information that can be gained by an eavesdropper {\it before} any PA and DR procedures are invoked. Why we focus on what happens before these procedures? These procedures are only necessary since QKD schemes usually do not {\it prevent} the eavesdropper from gaining information. Instead, QKD schemes are designed to enable Alice and Bob {\it detect} eavesdropping attempts (and abort if they find such attempts). We aim to reduce the amount of information accessible to an adversary in the first place (before the PA and DR procedures are employed), thereby increasing the capacity of the channel and diminishing the need for these expensive procedures. 

The game goes as follows. The participants in the game are Alice, Bob, and Eve. We assume that the participants can generate qubits in the computational basis, apply quantum gates to the qubits, and measure qubits. Alice and Bob are communicating in the game via a noiseless quantum channel and an authenticated classical public channel. Eve has full access to the quantum channel and is constantly listening to the public channel. Eve is computationally unbounded. At the first stage of the game, the parties are given a positive integer input $n$. Then, Alice and Bob engage in a QKD protocol of their choice to obtain a key of $\approx n$ bits (while not using PA or DR procedures). They are allowed to perform up to $\approx 4n$ transmissions of qubits between them, where half of the qubits are used for eavesdropping-check. As mentioned, Eve has full access to the quantum channel. When Alice and Bob invoke the QKD scheme, Eve deploys an attack strategy of her choice. She may intercept transmitted qubits, measure them, replace them with ancillary qubits of her choice, apply quantum gates to qubits, and perform arbitrary computations. At the last stage of the game, Alice and Bob and Eve decide if they want to abort the game. If they do, then the game is aborted and no one wins. If not, then all parties simultaneously announce their output. Alice outputs her key, an $n$-bit string $k_A=a_1\dots a_n$, Bob outputs his key, an $n$-bit string $k_B=b_1\dots b_n$, and Eve outputs a pair $(e,i)$, where $i$ is an integer and $e$ is a bit. Eve wins the game if she output a pair $(e,i)$ for which $a_i=b_i=e$. This is equivalent to Eve correctly guessing a key bit. (Below, we sometimes say that Eve outputs $\perp$ to indicate that she chose to abort.)

Observe that, Eve can always win the game with probability $\tfrac{1}{2}$ by picking the bit $e$ in random and setting, e.g., $i=1$. The advantage gained by a particular attack strategy over a particular QKD scheme is $|p_{success}-\tfrac{1}{2}|$, where $p_{success}$ is the probability that eve wins the game. 
\\ 

\noindent{\bf The suggested attack.} We now describe our WM attack against the BB84 scheme. Eve randomly picks $j\in\{1,\dots,4n\}$, prepares an ancilla $\ket{0}$ qubit, intercepts the $j$'th qubit transmitted from Alice to Bob, applies $\W$ to the intercepted qubit and the ancilla (the intercepted qubit is the control and the ancilla is the target), and sends Alice's qubit to Bob. Then, Eve measures the ancilla and obtains an outcome $e$ (the attack is illustrated in Figure 7).

\begin{center}
\includegraphics[scale=0.65]{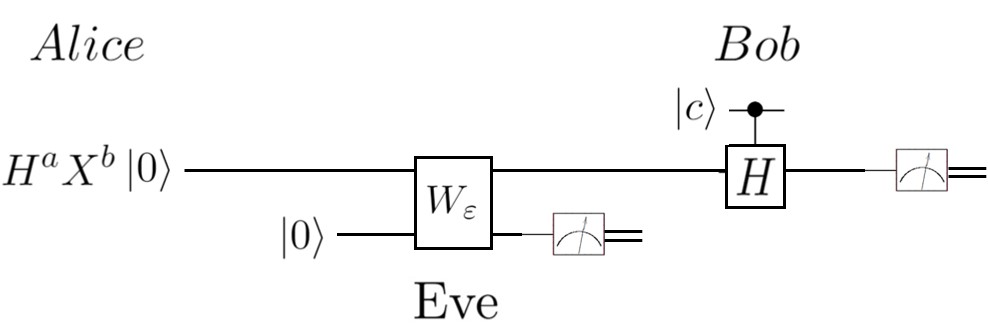}\\
\begin{minipage}{0.55\linewidth}
\small \textit {Figure 7: The weak measurement attack on BB84.} 
\end{minipage}
\end{center}

Next, Eve is listening to the discussion of Alice and Bob over the public channel and finds whether Bob measured the $j$'th qubit in the right basis (i.e, if $a=c$). If not, Eve's output is set to $\perp$ (i.e., she aborts). If $a=c$, then Eve keeps on listening to find whether the $j$'th qubit was used by Alice and Bob for eavesdrop-check or not. If it was, then Eve's output is set to $\perp$. If not, then Eve outputs $(e,i)$, where $i$ is the position of the $j$'th qubit after omitting the qubits for which $a\neq c$. Observe that, in this case, the outcome of Bob's measurement on the $j$'th qubit is Bob's $i$'th key-bit. We not that Alice and Bob abort only if they used the bit for eavesdropping-check and got different results. Hence, if Eve delivers an output (and do not abort) then Alice and Bob also do not abort.

We now analyze the attack with respect to the key bit guessing game -- what is the advantage of our WM strategy over the BB84 scheme? We are also interested in the following question -- what is the probability that Alice and Bob detect Eve's presence and abort? We are only interested in the cases where Alice and Bob measured the $j$'th qubit in the same basis, i.e., $a=c$. By Remark 2, we may assume it's always the case. To compute the WM attack's advantage, consider the system of two qubits where the first qubit is the qubit transmitted from Alice to Bob and the second qubit is the ancillary qubit used by Eve for the WM attack. If $a=c=0$, then that system of two qubits is in the state $$(1-b)\bigl(\sqrt{1-\varepsilon}+\sqrt{\varepsilon}\cdot i \bigr)\ket{00}+b\cdot \sqrt{1-\varepsilon}\cdot \ket{10} + \sqrt{\varepsilon}\cdot i \cdot b \cdot \ket{11},$$
and if $a=c=1$, then the system of two qubits is in the state $$\tfrac{\sqrt{1-\eps} +(-1)^b \sqrt{1-\eps}+i\cdot\sqrt{\eps}}{2}\ket{00}+(-1)^b\cdot \tfrac{i\cdot \sqrt{\eps}}{2} \ket{01}+ \tfrac{\sqrt{1-\eps}+i\cdot\sqrt{\eps}-(-1)^b\sqrt{1-\eps}}{2}\ket{10} -(-1)^b\cdot \tfrac{i\cdot\sqrt{\eps}}{2}\ket{11}.$$ We use the probabilities of the different possible outcomes of measurements of Bob and Eve given by these states to compute the total success probability of Eve given that $a=c$ (see Figure 8). 

The pairs $(x,y)$ in the bottom of the probabilities tree indicate the outcomes of the measurements of Bob ($x$) and Eve ($y$). The numbers in the green rectangles indicate the probabilities of the cases in which Eve correctly guessed the key-bit without causing an erroneous outcome for Bob (namely, $b=x=y$). This happens with probability $\tfrac{1}{4}+\tfrac{\eps}{4}+\tfrac{4-\eps}{16}+\tfrac{\eps}{16}=\tfrac{1}{2}+\tfrac{\eps}{8}$. The numbers in the red ovals indicate the probabilities of the cases in which Eve's attack resulted in Bob measuring an erroneous result ($x \neq b$). This happens with probability $4\cdot \tfrac{\eps}{16}=\tfrac{\eps}{4}$, and in these cases, if Alice and Bob use this bit for eavesdropping-check, then they will detect Eve's presence and abort. The purple hexagons indicate the probabilities of the cases in which Bob gets the right result, and Eve fails in guessing the key bit. In these cases, if Alice and Bob use this bit for eavesdropping-check, they will not detect Eve's presence. We conclude that using the WM-attack described above via the $\W$ gate, Eve can gain an $\tfrac{\eps}{8}$ advantage in guessing a key-bit while limiting the probability of getting caught to $\tfrac{\eps}{4}$.

\begin{center}
\includegraphics[scale=0.6]{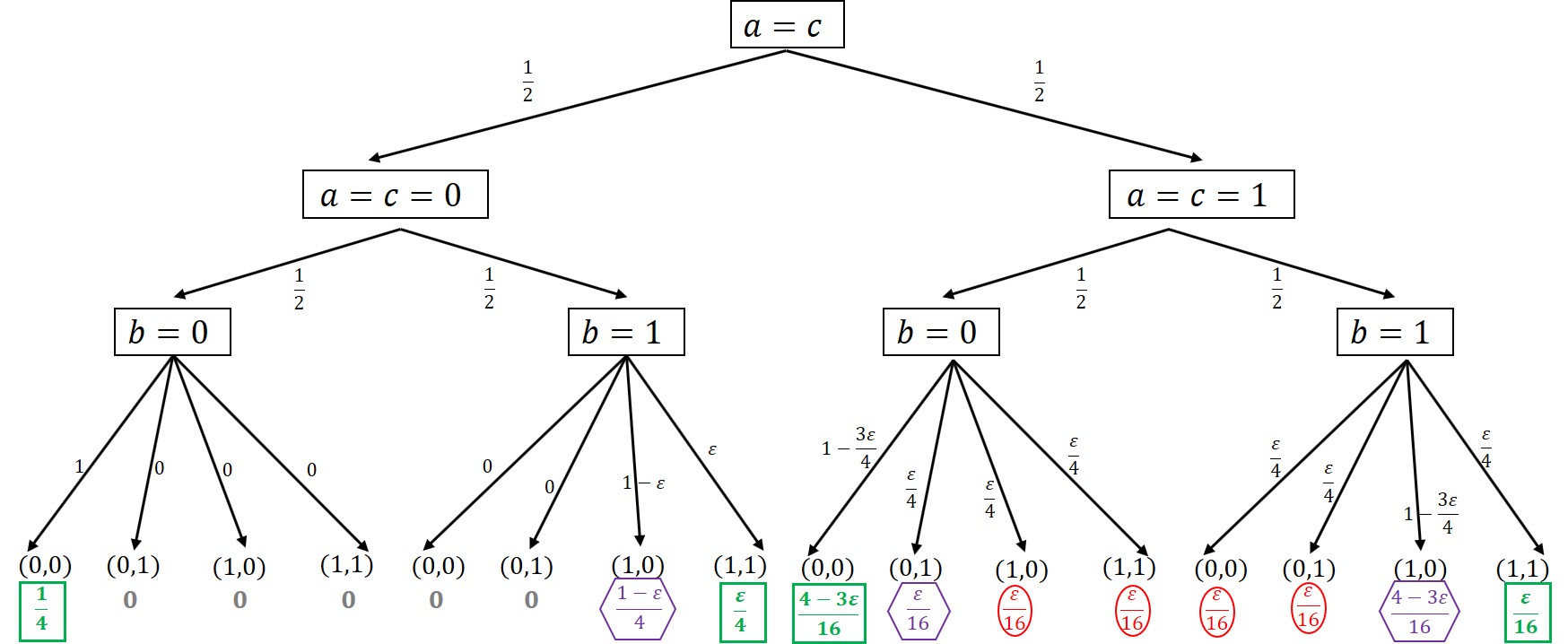}\\
\begin{minipage}{0.6\linewidth}
\small \textit {Figure 8: Probabilities of possible outcomes for the $\W$ attack.} 
\end{minipage}
\end{center}

\noindent{\bf A weak measurement attack on the DL04 scheme}. We now briefly review the QKD scheme suggested by Deng and Long in \cite{137} (hereafter, the DL04 scheme), suggest a WM-based attack for it, and analyze the attack's advantage with respect to the key bit guessing game. At the first stage of the DL04 scheme, Bob picks uniformly random bits $a$ and $b$ and generates the qubit $H^aX^b\ket{0}$. The bit $a$ encodes a choice of a basis (zero for computational, and one for Hadamard), and the bit $b$ is used for either eavesdropping check or generation of the key. Bob repeats the process (independently) $2n$ times and transmits the $2n$ qubits to Alice. Next, Alice randomly picks some of the qubits, say $n$, measures each of the selected qubits in either the standard or Hadamard basis (randomly) and announces the outcomes to Bob\footnote{This scheme can also be improved at this stage by using the same idea that we mentioned at Remark 2. Instead of Alice randomly choose the measurement basis, she can tell Bob which qubits she chose, Bob will reveal the corresponding $a$'s, and Alice will use this information to measure the qubits in the right basis. This will improve the probability of detecting possible adversarial eavesdropping attempts and the key generation rate.}. Next, if Alice and Bob find that the error rate is low enough (say, no errors were found), then there are $n$ qubits left (the ones that were not measured) with which they continue to the next stage. Now, Alice picks $c\in\{0,1\}$ and applies a $c$-conditioned $U$ gate to the first qubit, where $U=\begin{psmallmatrix} 0 & 1 \\ -1 & 0  \end{psmallmatrix}$. The unitary $U$ interchanges (up to a global phase factor) the elements of each of the relevant bases. I.e., $\ket{0}\xleftrightarrow[]U{\ket{1}}$ and $\ket{+}\xleftrightarrow[]U{\ket{-}}$. The bit $c$ is the transmitted key bit. Alice repeats the process for all the $n$ qubits she has not measured and sends the qubits back to Bob. Bob uses his knowledge of $a$ to decrypt the qubits by applying an $a$-conditioned Hadamard gate, and measures them to get $b\oplus c$, and uses his knowledge of $b$ to extract $c$. The scheme is illustrated in Figure 9.  

\begin{center}
\includegraphics[scale=0.6]{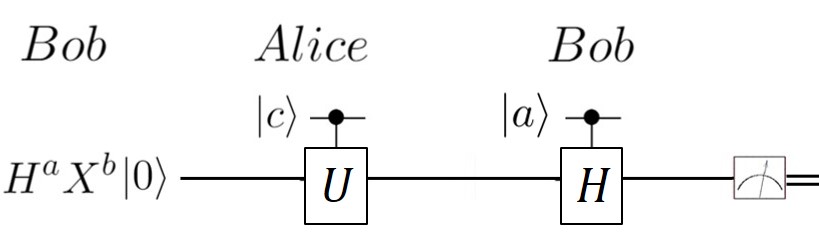}\\
\begin{minipage}{0.3\linewidth}
\small \textit {Figure 9: The DL04 scheme.} 
\end{minipage}
\end{center}

\noindent{\bf The suggested attack.} We now describe a WM-based attack on the DL04 scheme. This attack is based on the same idea as the attack on the BB84 scheme described above, and uses the $\W$ gate. The attack goes as follows. Eve randomly picks $j \in \{1,\dots,2n\}$. The $j$'th qubit is the objective qubit for the attack. Eve prepares an ancilla $\ket{0}$ qubit, intercepts the $j$'th qubit transmitted from Bob to Alice, applies $\W$ to the intercepted qubit and the ancilla (the intercepted qubit is the control and the ancilla is the target), and sends Bob's qubit to Alice. Eve measures the ancilla and obtains an outcome $e_1$. Next, Eve is listening to the measurement outcomes of Alice, announced over the public channel, and finds whether Alice measured the $j$'th qubit for eavesdropping check. If she did, Eve outputs $\perp$. If not, then Eve prepares another $\ket{0}$ ancilla. Denote by $i$ the new location of the objective qubit among the $n$ qubits that were not measured. Eve intercepts the $i$'th qubit transmitted from Alice to Bob, applies $\W$ to the qubit and the ancilla, and sends the qubit to Bob. Eve measures the (new) ancilla, obtains an outcome $e_2$, and outputs $(e_1\oplus e_2,i)$. The attack is illustrated in Figure 10.

\begin{center}
\includegraphics[scale=0.6]{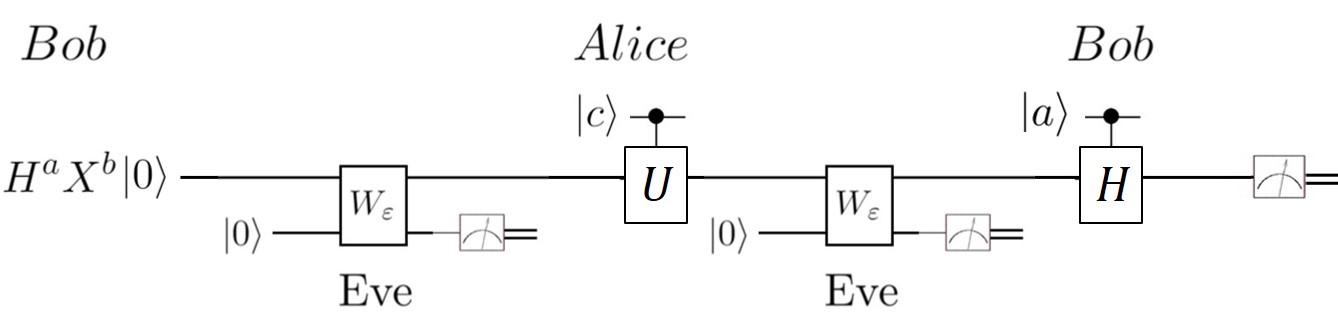}\\
\begin{minipage}{0.45\linewidth}
\small \textit {Figure 10: The WM attack on the DL04 scheme.} 
\end{minipage}
\end{center}

We now analyze the attack with respect to the key bit guessing game -- what is the advantage of our WM strategy over the DL04 scheme? Again, we are also interested in the probability that Alice and Bob detect Eve's presence and abort. We begin with some intuition. As before, if $a=0$, applying $\W$ does not change the objective qubit. In these cases, Alice and Bob do not detect Eve's presence, and Eve gains an $\mathcal{O}(\eps^2)$ advantage. The exponent $2$ comes from the fact that Eve must correctly detect the state in both directions of the transmission. When $a=1$, Eve gets no information but only slightly disturbs the state.

First, we consider the case in which Alice used the objective qubit (the one chosen by Eve), for eavesdropping check. In this case, Eve outputs $\perp$. What is the probability that Alice and Bob Detect Eve's presence on the line in these cases? Observe that this case (partly illustrated at the left part of Figure 8), is completely identical to the BB84 case described above (illustrated in Figure 5). Here, Bob is the one that generates one of the four qubits $\ket{0},\ket{1},\ket{+}$ or $\ket{-1}$ (with probability $\tfrac{1}{4}$ each), Eve's attack is identical (applying $\W$ with an ancilla), and Alice is the one who measures in the standard or Hadamard basis. As computed above, the probability that Alice and Bob disagree (and hence detect Eve's presence) is $\tfrac{\eps}{4}$. 

Next, we consider the case in which Alice did not choose the objective qubit for the eavesdropping check. Now, Alice applies a $c$-conditioned $U$ to the objective qubit and transmits it back to Bob. Again, Eve applies $\W$ to the objective qubit with an ancilla and measures the ancilla to obtain an outcome $e_2$. What is the probability that Eve's guess on $c$ is correct, i.e., $e_1 \oplus e_2 =c$? We have\\

\noindent {\bf Lemma 3.} {The probability that Eve's guess is correct is $\tfrac{1}{2}+\tfrac{6 \eps ^2 -3 \eps ^3}{16-8\eps}$.}

\noindent {\bf Proof of Lemma 3.}
\noindent {\it Proof.} We begin with some notations regarding the WM attack on DL04 described above. Recall that, at the first stage of DL04, Bob chooses $a,b\in\{0,1\}$ uniformly at random and sends $H^aX^b\ket{0}$ to Alice. We denote by $\ket{\psi_1}$ the two qubit system whose first qubit is the qubit sent from Bob to Alice, and the second qubit of $\ket{\psi_1}$ is Eve's (first) $\ket{0}$ ancilla. Namely, $\ket{\psi_1}= (H^aX^b\ket{0})\otimes\ket{0}$. $\ket{\psi_2}$ denotes $\W \ket{\psi_1}$. Recall that Eve measures the right qubit of $\ket{\psi_2}$ to obtain $e_1$. We denote by $\ket{\psi_3}$ the two-qubit system whose left qubit is the left qubit of $\ket{\psi_2}$ after Eve measures the right qubit of $\ket{\psi_2}$, and the right qubit of $\ket{\psi_3}$ is Eve's new $\ket{0}$ ancilla. $\ket{\psi_4}$ is the system obtained from $\ket{\psi_3}$ after Alice applies a $c$-conditioned $U$ to its left qubit. $\ket{\psi_5}$ denotes $\W\ket{\psi_4}$. These notations are illustrated at Figure 11.

\begin{center}
\includegraphics[scale=0.5]{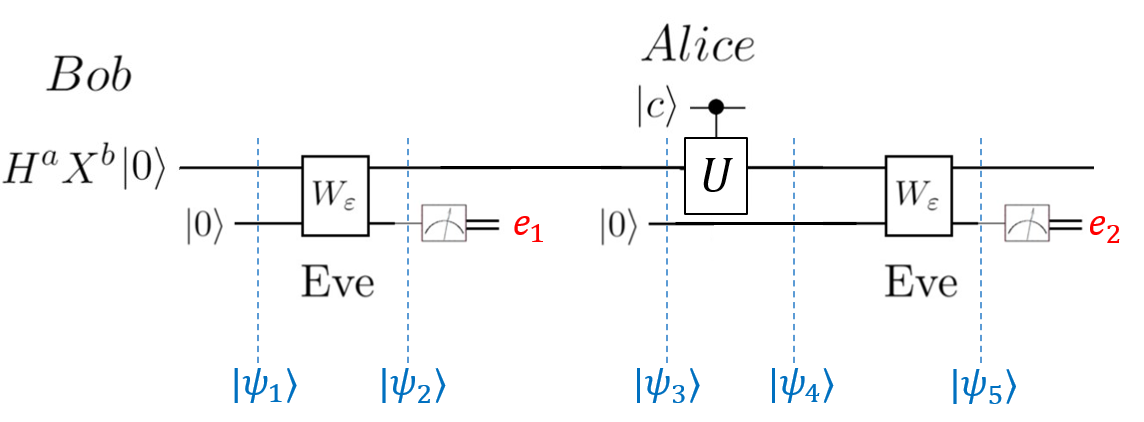}\\
\begin{minipage}{0.36\linewidth}
\small \textit {Figure 11: The WM attack on DL04.} 
\end{minipage}
\end{center}

Recall that Eve measures the right qubit of $\ket{\psi_5}$ to obtain $e_2$. Eve's guess is $e_1\oplus e_2$. The guess is correct if $e_1\oplus e_2=c$. To compute the probability of Eve guessing $c$ correctly we examine all the possibilities for $(a,b,c)\in \{0,1\}^3$. Each possibility occurs with probability $\tfrac{1}{8}$ (We assume that $c$ is chosen uniformly at random).\\

\begin{itemize}

    \item First case: $a=0$. 
    
    \begin{itemize}
        \item Assume $(b,c)=(0,0)$. In this case, $\ket{\psi_1}=\ket{00}$, and hence $\ket{\psi_2}=\W \ket{\psi_1}= \sqrt{\varepsilon} i  \cdot CNOT + \sqrt{1-\varepsilon} \cdot I\otimes I \ket{00} = \ket{00}.$ Next, Eve measures the right qubit of $\ket{\psi_2}$ and obtains $e_1=0$ with probability $1$. Now, $\ket{\psi_3}=\ket{00}$, and since $c=0$, we have $\ket{\psi_4}=\ket{00}$ and $\ket{\psi_5}=\ket{00}$ as well. Measuring the right qubit of $\ket{\psi_5}$ Eve obtains $e_2=0$ with probability $1$, which implies that Eve's guess in this case is $e_1 \oplus e_2 = 0 \oplus 0 =0$. Since here $c=0$, the guess is correct. This contributes $\tfrac{1}{8}$ to the total success probability. 
        \item Assume $(b,c)=(0,1)$. Here, $e_1$ and $\ket{\psi_3}$ are the same as in the previous case since Alice's choice of $c$ is only reflected at $\ket{\psi_4}$. Now, $\ket{\psi_4}=\ket{10}$, and hence $\ket{\psi_5}=\W \ket{10}= i\sqrt{\eps}\ket{11}+\sqrt{1-\eps}\ket{10}$. When Eve measures the right qubit of $\ket{\psi_5}$ she obtains the outcome $e_2=1$ with probability $\eps$, and the outcome $e_2=0$ is obtained with probability $1-\eps$. The former possibility implies a correct guess (since $e_1\oplus e_2 = 0 \oplus 1 =1 =c$), which contributes $\tfrac{\eps}{8}$ to the total success probability.
        \item Assume $(b,c)=(1,0)$. Here, $\ket{\psi_1}=\ket{10}$. Now, $\ket{\psi_2} =\W \ket{10}=i\sqrt{\eps}\ket{11}+\sqrt{1-\eps}\ket{10}$. Measuring the right qubit of $\ket{\psi_1}$ Eve obtains $e_1=1$ with probability $\eps$ and $e_1=0$ with probability $1-\eps$. Either way, $\ket{\psi_4} = \ket{10}$ and $\ket{\psi_5} =i\sqrt{\eps}\ket{11}+\sqrt{1-\eps}\ket{10}$. Measuring the right qubit of $\ket{\psi_5}$, Eve obtains $e_2=1$ with probability $\eps$ and $e_2=0$ with probability $1-\eps$. Since $c=0$, the correct guesses come from the cases where $(e_1,e_2)=(0,0)$ or $(e_1,e_2)=(1,1)$. The former possibility has probability of $(1-\eps)^2$, and the latter occurs with probability $\eps^2$. This contributes $\tfrac{(1-\eps)^2}{8}+\tfrac{\eps^2}{8} =\tfrac{1-2\eps+2\eps^2}{8}$ to the total success probability. 
        \item Assume $(b,c)=(1,1)$. Since Alice's choice of $c$ is only reflected at $\ket{\psi_4}$, the probabilities for $e_1$ are as in the previous case, and $\ket{\psi_3}=\ket{10}$. Here $c=1$, and hence $\ket{\psi_4}=\ket{00}$, which implies that $\ket{\psi_5}=\W \ket{00}=\ket{00}$. Measuring the right qubit of $\ket{\psi_5}$, Eve obtains the outcome $e_2=0$ with probability 1. If $e_1=1$ was obtained before, then Eve's guess is correct since $1\oplus 0 = 1$. Since the outcome $e_1=1$ is obtained with probability $\eps$ we conclude that this case contributes $\tfrac{\eps}{8}$ to the total success probability. \\
        
    \end{itemize}
    All in all, the contribution of the case $a=0$ to the total success probability is $$\tfrac{1}{8}+\tfrac{2\eps}{8} + \tfrac{(1-\eps)^2}{8} + \tfrac{\eps^2}{8}=\tfrac{1}{4}+\tfrac{\eps^2}{8}.$$\\

    \item Second case: $a=1$. First, we note that the left qubit of $\ket{\psi_1}$ is $H^aX^b\ket{0} =H\ket{b}=\ket{0}+(-1)^b\ket{1}$. If $b=0$ then the left qubit of $\ket{\psi_1}$ is $\ket{+}$, and if $b=1$ then it is $\ket{-}$. We write $\ket{\psi_1}=\ket{\pm}\ket{0}$. Now, $\ket{\psi_2}=\W \ket{\pm}\ket{0}= \W \Bigl(\tfrac{1}{\sqrt{2}}\ket{00} \pm \tfrac{1}{\sqrt{2}}\ket{10}\Bigr)$. Recall that $\W=\sqrt{\varepsilon} i  \cdot CNOT + \sqrt{1-\varepsilon} \cdot I\otimes I$. We have
    $$\ket{\psi_2}=\sqrt{1-\eps}\Bigl( \tfrac{1}{\sqrt{2}}\ket{00} \pm \tfrac{1}{\sqrt{2}}\ket{10}\Bigr)+i\sqrt{\eps} \Bigl( \tfrac{1}{\sqrt{2}}\ket{00} \pm \tfrac{1}{\sqrt{2}}\ket{11}\Bigr).$$
    Rearranging, we get
    $$ \ket{\psi_2}=\tfrac{\sqrt{1-\eps}+i\sqrt{\eps}}{\sqrt{2}} \ket{00} \pm \tfrac{\sqrt{1-\eps}}{\sqrt{2}}\ket{10} \pm i \tfrac{\sqrt{\eps}}{\sqrt{2}} \ket{11}.$$\\
    
    Measuring the right qubit of $\ket{\psi_2}$, Eve obtains the outcome $e_1=1$ with probability $\tfrac{\eps}{2}$, regardless of $b$. The outcome $e_1=0$ is obtained with probability $1-\tfrac{\eps}{2}$. We examine each possibility.\\
    
    \begin{itemize}
    
        \item If the outcome $e_1=1$ was obtained (which happens with probability $\tfrac{\eps}{2}$) then the left qubit of $\ket{\psi_2}$ collapses to $\ket{1}$. Now, there are two possibilities for $c$.
        \begin{itemize}
            \item If $c=1$ then $\ket{\psi_4}=\ket{00}$, which implies that $\ket{\psi_5}=\ket{00}$ as well. Measuring the right qubit of $\ket{\psi_5}$ Eve obtains the outcome $e_2=0$ with probability 1. Since we assume here $e_1=1$, Eve's guess, $1\oplus 0=1$, is correct. This contributes $\tfrac{\eps}{8}$ to the total success probability. 
            \item If $c=0$ then $\ket{\psi_4}=\ket{10}$, which implies that $\ket{\psi_5}=\W \ket{10}=i\sqrt{\eps}\ket{11}+\sqrt{1-\eps}\ket{10}.$ Measuring the right qubit of $\ket{\psi_5}$, Eve obtains the outcome $e_2=1$ with probability $\eps$, which compels a correct guess, $1\oplus 1=0=c$. This contributes $\tfrac{\eps^2}{8}$ to the total success probability.\\
            
        \end{itemize}
    
        \item If the outcome $e_1=0$ was obtained (which happens with probability $1-\tfrac{\eps}{2}$) then the left qubit of $\ket{\psi_2}$ collapses to 
        $$\frac{\tfrac{\sqrt{1-\eps}+i\sqrt{\eps}}{\sqrt{2}} \ket{0} \pm \tfrac{\sqrt{1-\eps}}{\sqrt{2}}\ket{1}}{\Big| \Big|\tfrac{\sqrt{1-\eps}+i\sqrt{\eps}}{\sqrt{2}} \ket{0} \pm \tfrac{\sqrt{1-\eps}}{\sqrt{2}}\ket{1}\Big| \Big|}.$$
        Using standard algebraic manipulations and joining Eve's (new) $\ket{0}$ ancilla we get
        $$ \ket{\psi_3}=  \sqrt{\tfrac{2}{2-\eps}} \Bigl(    \tfrac{\sqrt{1-\eps}+i\sqrt{\eps}}{\sqrt{2}} \ket{00} + \tfrac{\sqrt{1-\eps}}{\sqrt{2}}\ket{10}\Bigr).$$
        Now, there are two possibilities for $c$. 
        \begin{itemize}
            \item If $c=0$ then $\ket{\psi_4}=\ket{\psi_3}$. In this case, since $\ket{\psi_5}=\W\ket{\psi_4}$ we get

            $$\ket{\psi_5}=\tfrac{\sqrt{2-2\eps}}{\sqrt{2-\eps}} \Bigl( \tfrac{\sqrt{1-\eps}+i\sqrt{\eps}}{\sqrt{2}}\ket{00}+ \tfrac{\sqrt{1-\eps}}{\sqrt{2}} \ket{10} \Bigr) + \tfrac{i\sqrt{2\eps}}{\sqrt{2-\eps}} \Bigl( \tfrac{\sqrt{1-\eps}+i\sqrt{\eps}}{\sqrt{2}}\ket{00} +\tfrac{\sqrt{1-\eps}}{\sqrt{2}} \ket{11}      \Bigr)  .$$
            We rearrange $\ket{\psi_5}$ by the standard basis elements and see that the coefficient of $\ket{11}$ is $\tfrac{i\sqrt{2\eps}}{\sqrt{2-\eps}}\cdot \tfrac{\sqrt{1-\eps}}{\sqrt{2}}$and the coefficient of $\ket{01}$ in $\ket{\psi_5}$ is $0$. Hence, measuring the right qubit of $\ket{\psi_5}$ Eve obtains the outcome $e_2=1$ with probability $\alpha := \tfrac{\eps(1-\eps)}{2-\eps}$. Hence, the outcome $e_2=0$ is obtained with probability $1-\alpha$. If that happens, we have $e_1 \oplus e_2 = 0\oplus 0 = 0$, which yields a correct guess. This contributes $\tfrac{1}{4}(1-\tfrac{\eps}{2})(1-\alpha)$ to the total success probability. \\
            
            \item If $c=1$ then Alice applies $U$ to the left qubit of $\ket{\psi_3}$ and hence $$\ket{\psi_4}=  \sqrt{\tfrac{2}{2-\eps}} \Bigl(    \tfrac{\sqrt{1-\eps}+i\sqrt{\eps}}{\sqrt{2}} \ket{10} + \tfrac{\sqrt{1-\eps}}{\sqrt{2}}\ket{00}\Bigr).$$
            Now, $\ket{\psi_5}=\W\ket{\psi_4}.$ Here we get $$\ket{\psi_5}=\sqrt{\tfrac{2-2\eps}{2-\eps}} \Bigl(    \tfrac{\sqrt{1-\eps}+i\sqrt{\eps}}{\sqrt{2}} \ket{10}+ \tfrac{\sqrt{1-\eps}}{\sqrt{2}}\ket{00}\Bigr)+i\sqrt{\tfrac{2\eps}{2-\eps}} \Bigl( \tfrac{\sqrt{1-\eps} +i\sqrt{\eps}} {\sqrt{2}} \ket{11} +\tfrac{\sqrt{1-\eps}}{\sqrt{2}}\ket{00}\Bigr).$$
            Measuring the right qubit of $\ket{\psi_5}$ Eve obtains the outcome $e_2=1$ with probability $$\beta =\Bigl|  i\sqrt{\tfrac{2\eps}{2-\eps}} \cdot \tfrac{\sqrt{1-\eps} +i\sqrt{\eps}} {\sqrt{2}}  \Bigr|^2=\tfrac{\eps}{2-\eps}.$$ 
            In this case we get $e_1\oplus e_2 = 0 \oplus 1 =c$, which yields a correct guess. This contributes $\tfrac{\beta}{4}(1-\tfrac{\eps}{2}) $ to the total success probability. \\
        \end{itemize}
    \end{itemize}
    All in all, the contribution of the case $a=1$ to the total success probability is $$\tfrac{\eps}{8}+ \tfrac{\eps^2}{8}+ \tfrac{1}{4}(1-\tfrac{\eps}{2})(1-\alpha)+ \tfrac{\beta}{4}(1-\tfrac{\eps}{2}).$$
\end{itemize}
Using standard algebraic manipulations, the reader may readily verify that the contributions of the cases $a=0$ and $a=1$ add up to a total success probability of $\tfrac{1}{2}+ \tfrac{6\eps^2-3\eps^3}{8(2-\eps)}$.   $\square$

To conclude, using our WM attack on DL04, Eve gains an $\mathcal{O}(\eps^2)$ advantage in the key bit guessing game while being caught with probability $\tfrac{\eps}{4}$. \\ 

\noindent {\bf Our RBE-based WM-resilient CNOT QKD scheme.} {\color{black} Our RBE QHE scheme raises a QKD scheme that is resilient to WM-based attacks as suggested above. We now present the RBE-based $CNOT$-QKD scheme (also illustrated in Figure 12). We assume that $b=b_1\dots b_n \in \{0,1\}^n$ is a string of $n$ classical bits held by Alice. Alice wants to send $b$ to Bob with information-theoretic security. The scheme goes as follows.} \\

\begin{center}
\includegraphics[scale=0.6]{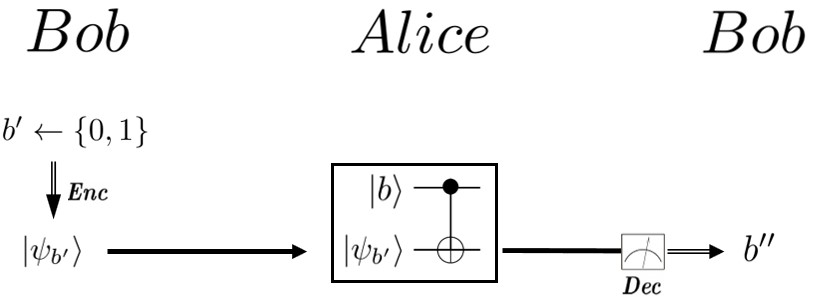}\\
\begin{minipage}{0.36\linewidth}
\small \textit {Figure 12: Sharing Key by Random Basis.} 
\end{minipage}
\end{center}

{\color{black} 
\fbox{%
    \parbox{\textwidth}{%
        \center {\it RBE-based Quantum Key Distribution.}
\begin{enumerate}
\item Bob randomly (uniformly) samples a $2n$-bit string from $\{0,1\}^{2n}$, denoted $b'=b_1' \dots b_{2n}'$. 
\item Bob RBE-encrypts each bit $b'_i$, $1\leq i \leq 2n$, independently, obtains $\ket{\psi_{b'_i}}$ and sends the encryptions to Alice.
\item Alice picks $n$ of the qubits at random and publishes the positions she chose. Then, Bob publishes the corresponding keys used to encrypt them, denoted $(\theta_i,\varphi_i)$.  
\item Alice RBE-decrypts the chosen qubits and publishes the outcomes. Then, Bob checks that the outcomes match the $b'_i$s sent. If they match, they proceed to the next stage. Otherwise, they abort due to a possible adversarial eavesdropping attempt. 
\item For each of the $n$ qubits that Alice did not measure at the previous stage, Alice does as follows.  If $b_i=1$, she applies a $NOT$ gate to the $i$'th qubit; otherwise, she leaves it unchanged. 
\item Finally, Alice transmits the $n$ qubits she did not measure to Bob. Bob decrypts these qubits to obtain a new string, $b''$. 

\item Bob computes the exclusive-or of $b''$ and the $n$-bit string obtained from $b'$ after removing the $n$ bits Alice chose at stage 3 (denoted $\tilde{b}$) and obtains $b$.\\
\end{enumerate}
    }%
}

\medskip

The security of the protocol follows from that of the RBE QHE scheme. While the BB84 and DL04 protocols are vulnerable to WM-based attacks, the RBE-based QKD protocol is robust to such attacks. The WM attacks we suggest take advantage of the fact that in the BB84 and DL04 protocol, the target qubit is $50\%$ of the time on the standard basis. In these cases, one may copy and measure it without disturbing it. Disturbance of the qubit (and the possibility of getting caught) happens when the qubit is in a non-standard basis. Additionally, an adversary can reduce the probability of getting caught by choosing a smaller $\eps$. But in the RBE-based QKD protocol, a qubit is on a non-standard basis $100\%$ of the time, leaving no place for such attacks. }

\noindent{\bf Remark 3.} We note that the DL04 scheme can be modified to obtain a {\it quantum secure direct communication} (QSDC) scheme. QSDC schemes are similar to QKD schemes. These schemes enable Alice and Bob to IT-securely exchange not only a random key, but also an arbitrary message of their choice (see, e.g., \cite{137,136} and the references therein). The WM-based attack against the DL04 scheme suggested above also works against the modified (QSDC) version of the DL04 scheme. Our RBE-based CNOT QKD scheme can also be modified to obtain a QSDC scheme. Unlike the DL04 QSDC scheme, our QSDC scheme is resilient to such WM-based attacks.

\section{Discussion}
\label{sec:4:5}
In this work, we have identified a useful family of orthonormal bases of $\hil$ and suggested a QHE encryption scheme of classical data based on this family -- the RBE QHE scheme. We have proved that our scheme is IT-secure, and discussed its homomorphic properties. The homomorphic operations that we support are supported by our scheme in an efficient, fully compact, non-interactive, perfectly correct, and IT-secure way and most importantly, with safer security in the face of adversarial attacks based on weak measurements. 

We have shown the usefulness of our QHE scheme for designing a protocol that enables two distant parties to obtain an entangled pair securely. In so doing, we first brought up the concept of securing the resource of entanglement. We also leveraged our RBE QHE scheme to design a QKD scheme. We suggested weak measurement based attacks on the BB84 and DL04 schemes to which our scheme is resilient. We note that, other known QKD schemes may also be attacked using similar WM-based constructions. Picking the encryption keys from an immense set of possible keys makes our scheme resilient to such weak measurement based attacks. One of the weaknesses of our scheme is that Alice must send Bob information about the angle of rotation over a secure channel. However, we note that the entangled resource pool can be built up front for the future, thus, allowing long setting time, including communication over secure channel. 
We believe that our new approach and techniques suggest a possible direction for future research on IT-secure quantum homomorphic encryption and quantum computation and information.
\\

\section{Data Availability Statement}
Data sharing not applicable to this article as no datasets were generated or analysed during the current study.

\bibliographystyle{alpha}
\bibliography{Main_Manuscript}

\end{document}